%% file: BrainParcellation.tex
\documentclass[3p,authoryear,nopreprintline]{elsarticle}
\input{format.tex}

\begin{document}

\begin{frontmatter}
\title{\textbf{Supervised brain node and network construction under voxel-level functional imaging}}

\author[1]{Wanwan Xu}
\author[1]{Selena Wang}
\author[3]{Chichun Tan}
\author[4]{Xilin Shen}
\author[4]{Wenjing Luo}
\author[4]{Todd Constable}
\author[2]{Tianxi Li}
\author[1]{Yize Zhao\footnote{Correspondence should be directed to: yize.zhao@yale.edu}.}
\address[1]{Department of Biostatistics, Yale School of Public Health, Yale University}
\address[3]{Department of Biostatistics, School of Public Health, Brown University}
\address[4]{Department of Radiology and Biomedical Imaging, Yale School of Medicine, Yale University}
\address[2]{School of Statistics, University of Minnesota}

\begin{abstract}
 Recent advancements in understanding the brain's functional organization related to behavior have been pivotal, particularly in the development of predictive models based on brain connectivity. Traditional methods in this domain often involve a two-step process by first constructing a connectivity matrix from predefined brain regions, and then linking these connections to behaviors or clinical outcomes. However, these approaches with unsupervised node partitions predict outcomes inefficiently with independently established connectivity.  In this paper, we introduce the Supervised Brain Parcellation (SBP), a brain node parcellation scheme informed by the downstream predictive task. With voxel-level functional time courses generated under resting-state or cognitive tasks as input, our approach clusters voxels into nodes in a manner that maximizes the correlation between inter-node connections and the behavioral outcome, while also accommodating intra-node homogeneity.  We rigorously evaluate the SBP approach using resting-state and task-based fMRI data from both the Adolescent Brain Cognitive Development (ABCD) study and the Human Connectome Project (HCP). Our analyses show that SBP significantly improves out-of-sample connectome-based predictive performance compared to conventional step-wise methods under various brain atlases. This advancement holds promise for enhancing our understanding of brain functional architectures with behavior and establishing more informative network neuromarkers for clinical applications.
\end{abstract}

\begin{keyword} 
Brain atlas \sep
Connectome-based predictive model \sep 
FMRI \sep 
Functional connectivity \sep
Spectral clustering \sep
Supervised learning
\end{keyword}
\end{frontmatter}

\section{Introduction}\label{sec:intro}
Understanding brain functional organization through large-scale networks and how such topology relates to behavior are two of the fundamental themes in neuroscience. By partitioning the brain into a collection of regions or nodes, whole-brain functional connectivity or connectome can be constructed using functional magnetic resonance imaging (fMRI) under a resting state or different cognitive tasks to characterize functional dependence, encapsulated within network configurations spanning across all the nodes. Subsequently, the established functional connectivity can serve as predictive entities for a set of learning methods called connectome-based predictive model (CPM) \citep{shenUsingConnectomebasedPredictive2017}, and have been linked with normal cognitive processes and a variety of disorders such as anxiety \citep{wang2021connectome}, obsessive-compulsive disorder \citep{wu2023connectome}, and Parkinson's disease \citep{wang2022antagonistic}. 


To establish functional connectivity, it is essential to introduce a brain parcellation scheme to define nodes, upon which functional connections are built.
Over recent decades, extensive efforts have been devoted to constructing brain parcellations that reflect neuroanatomical or functional organizations, offering insights into behavioral outcomes such as seizures, sclerosis lesions, cerebrovascular diseases, and other neurological disorders \citep{revell2022framework, shieeTopologypreservingApproachSegmentation2010, nowinski2013stroke, shenGroupwiseWholebrainParcellation2013}. Initial parcellation efforts created atlases based on brain cytoarchitecture or anatomical configurations, including the Brodmann-based automatic anatomic labeling atlas (AAL) \citep{brodmann1909vergleichende, amunts2015architectonic, tzourio2002automated}. However, such anatomical atlases often contained regions too coarse for effective functional connectivity studies. Alternatively, based on fMRI data, a growing number of brain parcellations have been defined under the assumption that functional signals within a node should be coherent. Some of these parcellations were generated under fine-scale functional connectivity by identifying homogeneous modules via diverse graphic models or boundary detection techniques \citep{craddock2012whole, shenGroupwiseWholebrainParcellation2013, fan2016human, thomas2011organization, gordon2016generation}. Others  \citep{ji2009parcellation,fan2021brain} directly worked with the raw time series and resorted to signal separation models including independent component analysis (ICA) to delineate functional alignment among the time series. Moving forward along this line, recent studies have indicated reconfigurations of brain functional architectures with sex, cognitive states, and task demands \citep{salehi2018exemplar, cole2014intrinsic, yeo2015functional}.  This growing body of research suggests that functional regions are dynamic in their boundary definitions, advocating for a flexible atlas that adapts to specific functional involvements.

\begin{figure}
\centering
\begin{subfigure}[b]{.39\textwidth}
  \centering
  \includegraphics[height = 10.5cm]{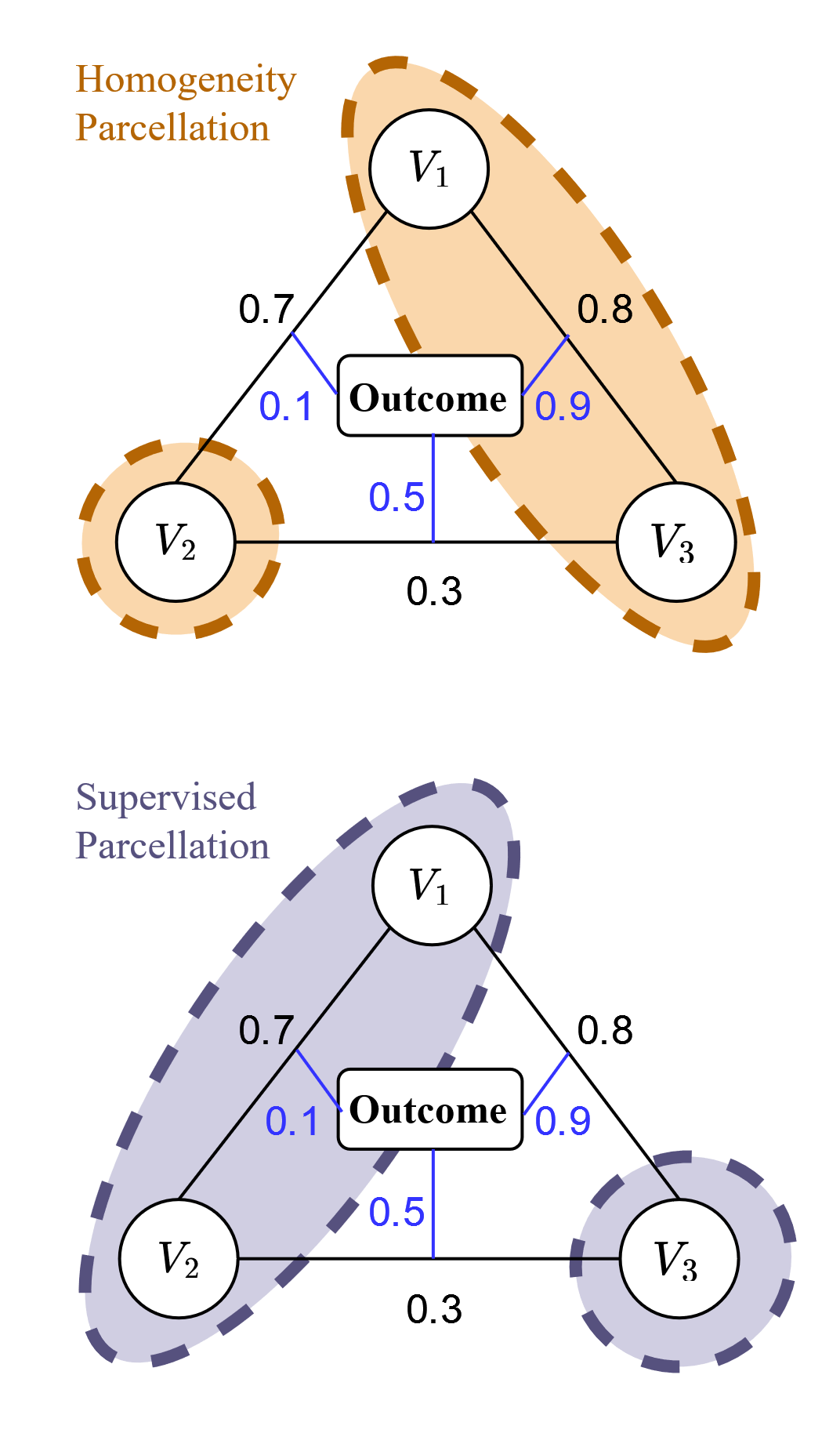}
  \caption{Synthetic example}
  \label{fig:illustration-toy}
\end{subfigure}
\begin{subfigure}[b]{.6\textwidth}
  \centering
  \includegraphics[height = 11cm]{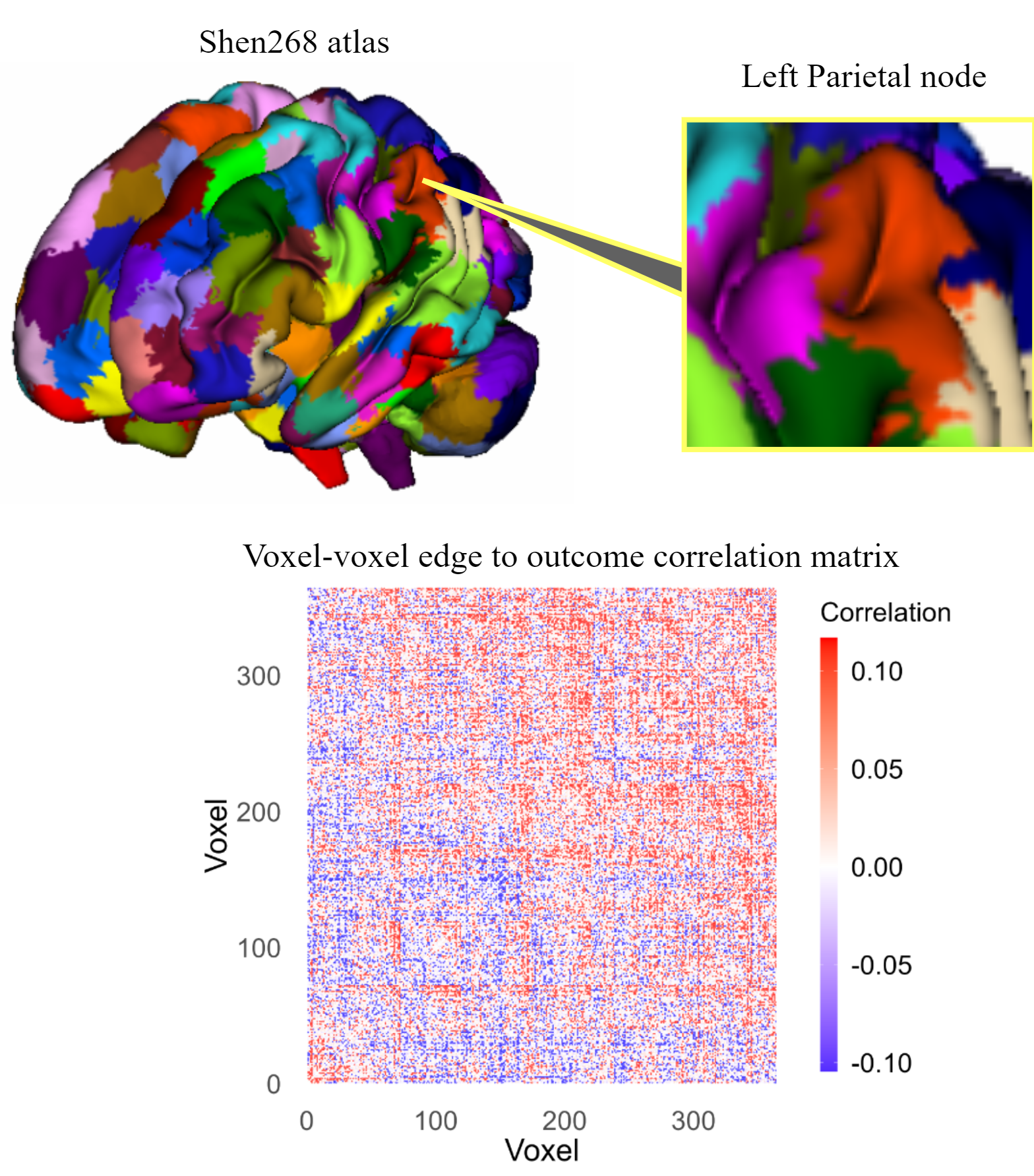}
  \caption{The $L.BA.39.1$ node from Shen268 atlas.}
  \label{fig:illustration-shen}
\end{subfigure}%
\caption{(a) Illustration of the proposed method using a synthetic example with three voxels labeled $V_1$, $V_2$, and $V_3$, represented by circles. Above black lines, numerical values indicate the strength of the functional connection between voxel pairs, while next to blue lines, values denote the strength of the association between each voxel-level connection and the target outcome. The top panel displays the existing regional homogeneity parcellation scheme with orange ellipses, and the bottom panel illustrates the proposed supervised parcellation with blue ellipses. (b) A detailed view of node 184 (left parietal area) from the Shen268 atlas \citep{shenGroupwiseWholebrainParcellation2013}, analyzed during the MID task from the ABCD study. Functional connections are quantified by the Pearson correlation of the time series, with only the top $25\%$ absolute values displayed in the heatmap. Positive correlations are shown in red, and negative correlations are in blue, providing a visual contrast for the cognitive composite score association.}
\label{fig:illustration}
\end{figure}

Conversely, in efforts to establish brain-to-behavior correspondence and develop predictive models under connectivity data, existing unsupervised parcellation schemes that aim for a marginal uniformity within individual regions may lose their relevance and can even become counter-productive. This is because when computing a functional connection between two nodes, the initial step involves averaging all voxel-level time series within a node. Subsequently, the statistical dependency is calculated between each pair of the averaged time courses. In this process, all the connectivity signals within a node are fully saturated, overlooking the potential intra-region functional heterogeneity linked with the target outcome \citep{luo2022inside}. To elucidate this concept, Figure \ref{fig:illustration-toy} presents a synthetic example with three voxels, $V_1, V_2, V_3$, represented as circles. Assuming that the functional edges between each pair of voxels are 0.8, 0.7, and 0.3, as illustrated, and that their associations with the outcome are 0.9, 0.1, and 0.5, respectively. Given the high similarity in voxel-wise functional signals between $V_1$ and $V_3$ indicated by a large correlation, it is expected that these two voxels would be grouped into the same node, separate from $V_2$ under an existing  homogeneity-focused parcellation scheme. However, this overlooks the highly predictive edge between them, leading to an average node-wise connectivity-to-outcome correlation of only 0.3.  In contrast, when the downstream learning objective is considered during the parcellation learning, node definitions should ideally reveal the strongest predictive edges. As depicted in the lower panel of Figure \ref{fig:illustration-toy}, a supervised parcellation scheme would group  $V_1$ and $V_2$ by keeping the highly predictive edges as inter-node ones. Such a parcellation yields an average correlation of 0.7 between node-wise functional connections and the outcome, demonstrating its potential effectiveness.

To this end, we introduce a novel Supervised Brain Parcellation (SBP) scheme that leverages the relationship between state-specific functional connectivity alterations and behavioral trait variations. We posit that the delineation of functional boundaries for node construction should not only reflect anatomical structure but also enhance the predictive power of network architectures and reliably identify functional neuromarkers. Our proposed supervised parcellation learning framework, inspired by regularized spectral clustering, takes into account the correspondence between the defined functional connections and the behavior outcome. In contrast to the existing atlases, SBP identifies brain nodes and networks that more accurately inform behavioral outcomes while preserving the brain spatial continuity. This trait-specific approach is considerably more adaptable to downstream learning tasks, and it operates independently in both training and testing phases. Through the analyses of two landmark studies, the Adolescent Brain Cognitive Development (ABCD) study and the Human Connectome Project (HCP) under their resting-state and task-based fMRI data, we demonstrate that voxel-to-node memberships are reconfigured under different cognitive conditions with the proposed supervised parcellations substantially enhancing the out-of-sample predictive performance for general intelligence compared with the existing parcellation methods. This work contributes to the evolving paradigm of brain parcel constructions by acknowledging that functional brain regions are not confined to static boundaries but could be considered with adaptability and responsiveness to their functional engagement. It illuminates the path toward objective-enhanced and information-supervised parcellations, with the resulting functional organizations providing significant benefits for establishing brain-to-behavior learning models and uncovering neurobiological network signals. 

The remaining of this paper is organized as follows: In Section \ref{sec:methods}, we introduce our experimental studies, provide empirical evidence on signal obscuration, detail our supervised parcellation learning framework, and describe the optimization algorithm and model evaluation procedures. In Section \ref{sec:results}, we apply the model to create parcellations using fMRI under varying cognitive conditions and behavioral data from the ABCD and HCP, and conduct independent validation and testing to evaluate predictive and replicative efficacy. The paper concludes with a discussion in Section \ref{sec:discussion}.

\section{Material and methods}\label{sec:methods}
\subsection{Data}\label{sec:data}
To demonstrate our method and show the power of the developed functional parcellations, we utilize resting-state and task-based fMRI and behavioral data collected for each subject from the ABCD and HCP studies. We also demonstrate the motivation for this work through a data example extracted from the ABCD study and illustrate the core idea of our method with a simulated dataset.

\subsubsection{ABCD study}\label{sec:abcd}
The ABCD study is an ongoing prospective study launched in 2015 to investigate brain development and adolescent health for more than 10,000 children aged 9 to 10 years from 21 sites across the United States \citep{garavan2018recruiting}. The study has been collecting a wealth of measures on brain imaging, biospecimens, and cognitive development measurements to support different dimensions of brain-to-behavior studies. For the fMRI images, each participant went through a scan session in a fixed order beginning with a localizer, acquisition of 3D T-weighted images, 2 runs of resting-state fMRI, T2-weighted images, 1-2 runs of resting-state fMRI and task-based fMRI.
 The ABCD study collected three task-based fMRI including an emotional version of the n-back task (nback) \citep{cohen2016impact}), the Monetary Incentive Delay (MID) task \citep{knutson2000fmri}, and the Stop Signal task (SST) \citep{logan1994spatial}. These task domains involve cognitive functions related to working memory, emotion regulation, reward processing, motivation, impulsivity, and impulse control \citep{casey2018adolescent}.  More details on the imaging acquisition across different sites and pre-processing are described elsewhere by \cite{greeneTaskinducedBrainState2018,casey2018adolescent,hagler2019image}.


We focused on the first release of fMRI data (Released 3.0.1). Raw DICOM images for \( 5,772 \) subjects were collected via ABCD fast track \citep{casey2018adolescent} and preprocessed using BioImageSuite \citep{papademetris2006bioimage}. All fMRI images were realigned to correct for motion and registered to the standardized \( 3mm \times 3mm \times 3mm \) common space. Only subjects with low-motion fMRI data (mean frame-to-frame displacement \( < 0.15mm \) and maximum frame-to-frame displacement \( < 2mm \)) at \( 3mm \) resolution were included in our analyses. We are interested in the total cognition composite score from the NIH Toolbox Cognition Function Battery \citep{akshoomoff2013,Heaton2014} as the target outcome. This score encapsulates two cognitive abilities: ``crystallized," based on past learning, and ``fluid," indicative of novel learning and information processing in unfamiliar contexts. We employed the total cognition composite score for its demonstrated test-retest reliability and its high correlation with cognitive summary scores in the literature \citep{Heaton2014}.
After preprocessing, \( 1589 \) subjects, consisting of \( 858 \) (\(54\% \)) female and \( 731 \) (\(46\% \)) male, were included in the study. The input images have dimensions of \( 61 \times 73 \times 61 \times 396 \), where the first three axes represent voxel coordinates, and the last represents time points. To enable comparisons with existing atlases, only \( 54,971 \) voxels included in the AAL atlas \citep{rollsAutomatedAnatomicalLabelling2020} were retained. Voxels with only zero time courses were treated as missing data. We computed the missing rate for each subject under each functional condition (Rest, MID, nBack) and excluded the top \( 5\% \) of subjects with the most missing voxels. For the remaining \( 1509 \) subjects, the overlapping voxels with complete time courses were included in the model. 
Finally, we constructed the voxel-level connectivity maps for each subject under each condition using Pearson correlation coefficients for the time courses between each pair of voxels.

\subsubsection{HCP}\label{sec:hcp}
We also independently evaluated our method using the HCP study to show the robustness of the predictive performance. The HCP aims to map macroscopic human brain circuits and their behavioral correlates in a large population of healthy adults. We utilized a subset of the HCP S900 release with subjects who have collected voxel-level fMRI data for all nine functional sessions, which included two resting-states and seven task-based scans. Subjects with excessive head motion, defined as a mean frame-to-frame displacement exceeding \( 0.1mm \) or a maximum frame-to-frame displacement exceeding \( 0.15mm \), were excluded from the analysis. Eventually, the analyses focused on a dataset of \( 494 \) subjects. The preprocessing followed the same procedures as \cite{luo2021within} and \cite{salehi2020there}, using the HCP minimal preprocessing pipeline \citep{glasser2013minimal} and subsequent processing with BioimageSuite \citep{joshi2011unified}.
Consistent with the ABCD study, we limited the analysis to the \( 54,971 \) voxels within the AAL atlas. 
After removing missingness, the common voxels with complete time courses across the \( 469 \) subjects were used. Similarly, we summarized a voxel-level functional connectivity matrix for each subject and each state as our input. For the behavior outcome, we consider the fluid intelligence score assessed using a form of Raven’s progressive matrices with 24 items \citep{bilker2012development}.

\subsection{Signal obscuration}
We first illustrate how current brain atlases could obscure crucial intra-node signals in functional connectivity analysis. Using data extracted from the ABCD study as an example, we focus on functional connectivity constructed under the MID task with nodes defined by a commonly used Shen atlas \citep{shenGroupwiseWholebrainParcellation2013} with 268 nodes. Particularly, we take the node 184 located in the left parietal area (Brodmann area 39) 
as a case study, exemplified in Figure (\ref{fig:illustration-shen}). Within this node, there are $364$ voxels, which together form $66,066$ intra-node voxel-level functional edges. When correlating each functional edge with the cognition composite score, we obtain a wide range of correlation strengths, from \(0.80 \times 10^{-6}\) to \(0.12\), with \(44.42\%\) showing negative correlations and \(55.58\%\) positive. The diverse range of values within a single node indicates the heterogeneity in how different parts of the node relate to cognitive behavior; and we can also observe these spreading signals in the heatmap represented in Figure (\ref{fig:illustration-shen}).  When further examining functional connections between node 184 and other nodes, the average significant correlation between functional connections and the cognition composite score is only 0.01, which is considerably lower than what might be expected from a more refined intra-node connectivity. This highlights an opportunity to enhance functional network predictive accuracy with a more informative parcellation scheme.

\subsection{Supervised brain parcellation}
Suppose that under a cognitive state, we have fMRI scans collected for $p$ voxels across the whole brain for each of the $n$ subjects. The behavior outcome of interest is measured and denoted as $\mathbf{o} = (o^{(1)},o^{(2)},\ldots,o^{(n)}) \in \mathbb{R}^{n}$ with $o^{(i)}$ representing the outcome for subject $i$. The general objective is to partition the \( p \) voxels into \( K \) groups, denoted by the voxel index sets $\{C_k\}_{k=1}^K$; and we require \( C_k \cap C_{k'} = \emptyset \) for \( k \neq k' \) and \( \cup_{k=1}^K C_k = \{1,2,\ldots,p\} \). By treating each group as a region/node, we can then establish node-level functional connectivity as neuromarkers to predict  behavior outcomes.

One way to tackle this brain partition problem is through graph community detection by constructing a weighted indirect graph $\mathcal{G}=(V, E, W)$. The vertex set $V$ consists of all the voxels, edge set $E$ includes all connections under weights $W$. We consider the statistical dependence of the functional time courses as the weight metric, reflecting the functional organization. The graph $\mathcal{G}$ can be uniquely represented by an adjacency matrix $A^{(i)}=(a^{(i)}_{jl})\in \mathbb{R}^{p\times p}$ for subject $i$, which also serves as a voxel-level functional connectivity matrix. Our objective here is to find a brain parcellation $\cup_{k=1}^K C_k$ that is shared among subjects. Despite some attempts at an individual brain parcellation \citep{salehi2020individualized}, our study aims for a parcellation that can be directly generalized to independent samples for predicting behavioral outcomes, making a groupwise parcellation structure ideal. To aggregate adjacency matrices $\{A^{(i)}\}_{i=1}^n$ across samples and form a groupwise adjacency matrix $A$, we consider the following options:
\begin{align}
   1)~ A = \frac{1}{n} \sum_{i=1}^n A^{(i)}\odot A^{(i)}; \quad 2)~ A = \frac{1}{n} \sum_{i=1}^n {A^{(i)}}; \quad 3)~ A = \frac{1}{n}\left\{\sum_{i=1}^n A^{(i)}\odot A^{(i)} - D^{(i)}\right\}.
\label{eq:averageA}
\end{align}
Here, $\odot$ represents the Hadamard product, and $D^{(i)}$ is the diagonal matrix called degree matrix with the row sum of $A^{(i)}$ as diagonal elements. Among those options, the first one is supported by rigorous statistical justifications that topological structures could be incorporated in the mean squared connectivity matrix \citep{lei2022bias}; the second one represents a commonly adopted sample mean matrix for a group; and the final one is a refined version for option one with rescaling to stabilize the algorithm. In our numerical studies, all three options are considered during implementation, and we observe similar performance among these realizations indicating a robustness of our numerical operations for the group-level adjacency information among voxels.

\begin{figure}[H]
    \centering
    \includegraphics[width = .9\linewidth]{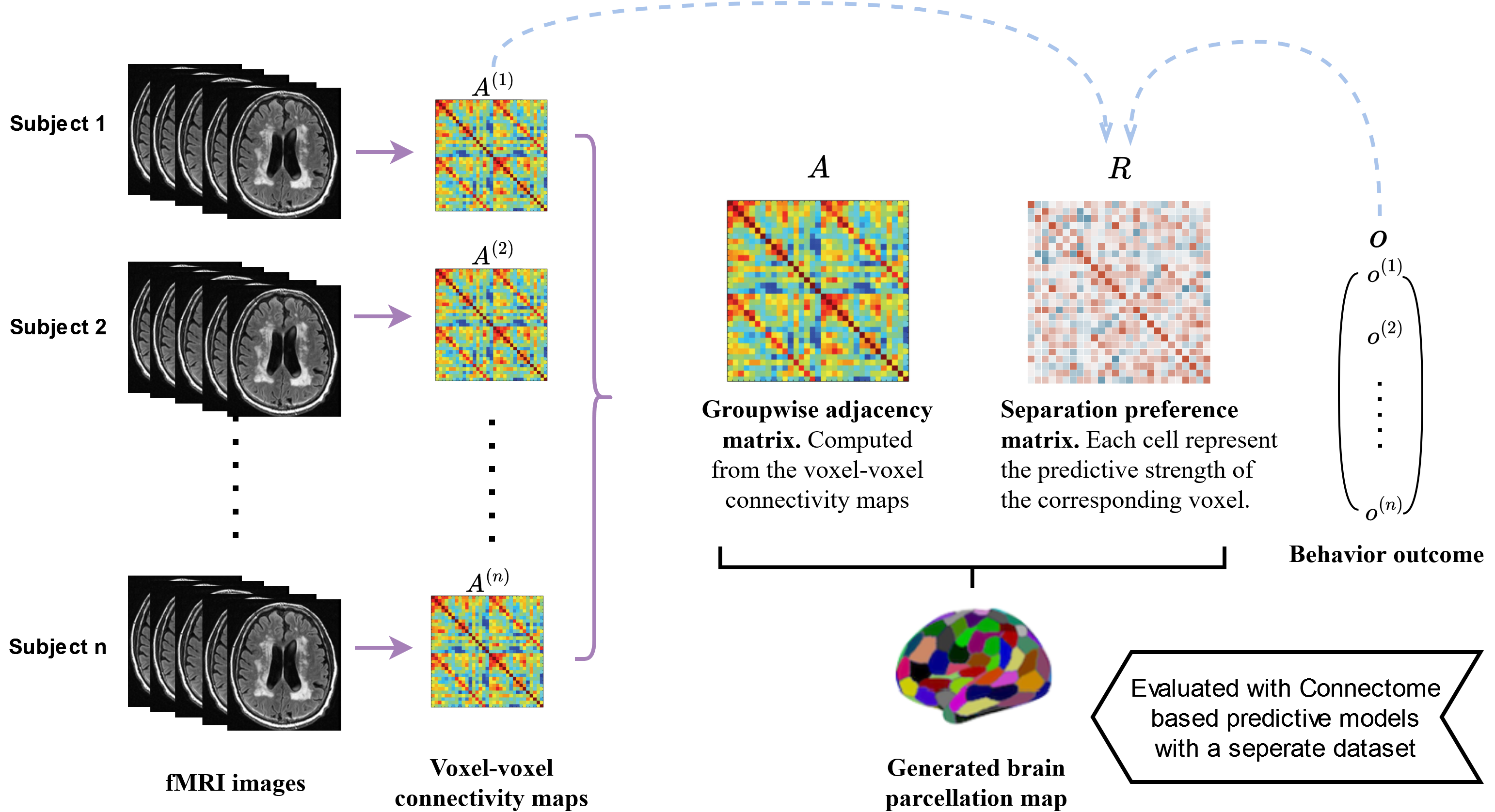}
    \caption{Flowchart of the proposed Supervised Brain Parcellation (SBP) scheme with both fMRI images and a behavior outcome.}
    \label{fig:flowchart}
\end{figure}

If we only focus on $A$ to perform community detection without consideration of the downstream predictive task, canonical spectral clustering is a powerful approach and has been widely used for constructing unsupervised brain parcellations \citep{KIM20102375,shenGroupwiseWholebrainParcellation2013}. This method typically utilizes the top eigenvectors of either network adjacency matrix or Laplacian matrix to segment the graph into distinct communities, leveraging spectral characteristics \citep{lei2022bias}. Through such a learning procedure, we ensure that the connections between different communities exhibit low similarity, whereas those within the same community demonstrate high similarity. To further incorporate correspondence with the behavior outcome of interest when constructing functional nodes,  on top of the adjacency matrix $A$, we create a separation preference matrix $R=(r_{jl}) \in \mathbb{R}^{p \times p}$ with the outcome information involved as a ``guidance" to learn the boundary of partitions.  The goal of the constructed preference matrix is to leverage the predictive power of the functional connection between voxel $j$ and voxel $l$ to regularize the spectral learning procedure. For each entry $r_{jl}$ in the preference matrix, we use the Pearson correlation between $\{a_{jl}^{(i)}\}_{i=1}^n$ and $\{o^{(i)}\}_{i=1}^n$ as a natural choice to characterize the association between the edge and the outcome. One can also use the inverse  pvalue of the Pearson correlation to regularize under statistical significance.  Combining two sides of information, the proposed SBP can be formulated as the following optimization procedure,
\begin{align}
\begin{split}
\min_{\{C_k,\mu_k\}_{k=1}^K} \quad & \sum_{k=1}^K \left(\sum_{j \in C_k} \|U_{j\cdot} - \mu_k\|^2  + \lambda \sum_{j,l \in C_k} r_{jl} \right), \\
\text{such that} \quad & C_k \subset [p],\quad C_k \cap C_{k'} = \emptyset, \quad \bigcup_{k=1}^K C_k = [p].
\end{split}
\label{eq:obj1}
\end{align}
Here $U_{j\cdot}$  denotes the $j$-th row of matrix $U \in \mathbb{R}^{p\times K}$ with $j=1,\dots,p$, where columns of $U$ are the $K$ leading eigenvectors of $A$ corresponding to the highest absolute eigenvalues; and $\mu_k = \sum_{j \in C_k} U_{j\cdot}/|C_k|$ taking the mean across rows $U_{j\cdot}$ for each $j \in C_k$ standardized by the cardinality of set $C_k$ denoted as $|C_k|$. Optimization \eqref{eq:obj1} indicates that our main loss function consists of two components.
The first component uses the groupwise functional connectivity adjacency matrix to guide the clustering through spectral learning. This part attempts to assign each voxel to the nearest centroid with the smallest Euclidean distance, which respects the pattern of the sample connectivity maps. The second component incorporates the influence of behavioral outcomes to refine the learning process. By introducing the regularization matrix $R$, we impose penalization to each of the entries $\{r_{jl}\}$ with $j, l$ belonging to the same node. By using the edge-to-outcome correlation values as entries in this matrix, we effectively regularize the stronger connections during the minimization process. This strategy is the key to ensure that the most informative signals are not lost. It leads to the clustering of voxels with highly predictive connections into distinct nodes, thereby preserving the integrity and predictive power of important functional connections. Within the optimization, the second component is controlled by a tuning parameter $\lambda$. If $\lambda = 0$, the model reduces to a standard spectral clustering of adjacency matrix $A$, as described in \cite{lei2022bias}.  To determine the optimal value  $\lambda$, we employ cross-validation method. The detailed procedure and considerations are explained in Section \ref{sec:hyper}. In essence, our learning strategy encompasses two key aspects: the similarity of connectivity patterns within nodes and the effective distribution of voxels. The latter involves grouping voxels linked by highly predictive connections into different nodes to preserve their predictive power. In Figure \ref{fig:flowchart}, we demonstrate a brief workflow of the proposed SBP.

\subsubsection{Coordinate descent algorithm on solving SBP}
Given our optimization problem is non-convex, we employ an efficient coordinate descent algorithm to conduct the minimization for \eqref{eq:obj1}. This method involves iterative updates on both voxel assignments and cluster mean values. The process begins by initializing the parcellation set $\{C_k\}$. In practice, we could start with the results from standard K-means clustering on the rows of matrix $U$. In each iteration, for voxel $j \in \{1,\ldots,p\}$, the new cluster label $k^j$ is determined as the one that minimizes the following objective function:
$$k^j = \mbox{argmin}_{k=1,\ldots,K} \Big\{-2U_{j\cdot}^T\mu_k + \|\mu_k\|^2  + \lambda \sum_{l \in C_k} r_{jl}\Big\}.$$
This function considers both the distance of the voxel to the cluster centroids and the separation preference matrix to jointly guide the cluster updates. Subsequently, the centroid of each cluster is recalculated as the mean of rows in $U$ corresponding to the newly assigned group labels in that cluster. We then perform a convergence check, assessing whether the change in the loss function is below a set tolerance level ($10^{-5}$ in our case). In Algorithm 1, we summarize each step in the SBP algorithm, which can be seen as a regularized adaptation of Lloyd's algorithm \citep{sabin1986global}. Of note, similar to the standard K-means problem, our objective is non-convex, and there is no guarantee of convergence to the global optimum. To mitigate this, we suggest multiple runs of the algorithm with varied initializations. The iteration yielding the smallest objective value is chosen as the final solution. In our numerical studies, we extensively evaluate the algorithm's performance and generally observe its effectiveness in converging towards a global solution.


\vspace{+4mm}
\fbox{\begin{minipage}{44em}
\singlespacing
\vspace{-3mm}
\begin{center}
    \textbf{Algorithm 1: Supervised Brain Parcellation (SBP)}

\end{center}
\textbf{Input:} 
\begin{itemize}[itemsep=-2pt]
\item[-]$\{A^{(i)}\}_{i=1}^n$: voxel-level connectivity matrices for $n$ subjects,
\item[-] $R$: separation preference matrix, 
\item[-] $K$: number of nodes,
\item[-] $\lambda$: tuning parameter.
\end{itemize}
\textbf{Procedure:}
\begin{enumerate}[itemsep=-2pt]
    \item Construct groupwise adjacency matrix $A$. 
    \item Perform eigendecomposition on $A$. Select the leading $K$ eigenvectors based on absolute eigenvalues. Let $U$ be the matrix with these eigenvectors as columns.
    \item Initialize the cluster assignments $C_1,\ldots, C_K$. Compute cluster cardinality $|C_k|$ and centroid $\mu_k = \sum_{j \in C_k} U_{j.} /|C_k|$ for each $k = 1,\ldots, K$.
    \item Repeat the following steps until converged or max iterations are reached:
    \begin{enumerate}[itemsep=-1pt,topsep =-1pt]
        \item Update voxel assignment: 
        
        For each voxel $j = 1,2,\ldots,p$, compute the loss for assigning it to the $k$-th node: $\|U_{j\cdot} - \mu_k\|^2  + \lambda \sum_{l \in C_k} R_{jl}$. Find the node $k^j$ with the smallest loss. Assign voxel $j$ to node $C_{k^j}$. 

        \item Update node centroid: 

        For each node $k = 1,2,\ldots,K$, compute the number of voxels assigned $|C_k|$, and update the $k$-th centroid to the corresponding mean of rows  $\mu_k = \sum_{i \in C_k} U_{i.} /|C_k|$. 

        \item Check for convergence: 
        
        If the centroids do not change or the change is smaller than the tolerance level, convergence is reached, exit loop. 
               
    \end{enumerate}
\end{enumerate}
\textbf{Output:} 
\begin{itemize}[itemsep=-2pt]
\item[-] $\{C_k\}_{k=1}^K$: assigned index sets, $k=1,2,\ldots, K$; 
\item[-] $\{\mu_k\}_{k=1}^K$: node centroids,  $k=1,2,\ldots, K$. 
\end{itemize}
\end{minipage}
}

\subsubsection{Hyperparameter selection}\label{sec:hyper}
To implement the algorithm to construct SBP, we consider a range of node numbers $K$ in our numerical studies, varying from 100 to 500 in light of most of the existing brain parcellation sizes. Consistent with previous studies \citep{shenGroupwiseWholebrainParcellation2013,salehi2018exemplar}, our goal is not to identify the optimal number of nodes for SBP as such an optimum may not exist. Instead, we focus on evaluating the model performance across various configurations of parcellation sizes.    
Regarding the tuning parameter $\lambda$ which balances the contribution to the parcellation construction between the spectral configuration of brain functional organizations and their induced predictive effect, we determine its value through a process involving training, validation and testing.  
For each dataset $(A^{(i)}, o^{(i)}, i=1,\dots,n)$, we randomly divide it into 10 folds with one-fold set aside  as a testing set, while the remaining 90\% samples are further randomly split into 10 folds for a cross-validation to determine $\lambda$. Under each split, 9 folds serve as training and 1 fold as validation. With a fixed $\lambda$, the SBP is trained on each training set to determine the node assignment $\{C_k\}_{k=1}^K$ to form a brain parcellation. The generated parcellation is then applied to the validation set to construct the regional functional connectivity based on the averaged node-level functional time course. Subsequently, a connectome-based predictive model (CPM) \citep{shenUsingConnectomebasedPredictive2017} is utilized to link the constructed connectivity matrix with the behavior outcome under this validation set with the predictive R-square obtained. The validation prediction under the current parcellation is then summarized by averaging the R-square across all 10 validation sets. Finally, we apply the parcellation with the optimal $\lambda$ from the validation to the testing set to construct regional functional connectivity. The predictive power is then assessed by CPM with the testing set R-square recorded. By repeating the above steps for each fold as a test set, we obtain the final averaged out-of-sample R-square. Throughout the procedure,  we maintain independence between the construction of the parcellation and the final predictive evaluation by separating the tuning parameter's determination from the testing samples. This ensures that the testing phase remains unbiased and the results are reliable.


\subsection{Reproducibility}\label{sec:reprod}
It is essential to ensure the reliability of our constructed SBP scheme by evaluating its reproducibility under each study. Similar to \cite{shenGroupwiseWholebrainParcellation2013}, we also adopt the widely used Dice's coefficient to quantify the similarity between two brain atlases that are generated from different runs of the model. For any two index sets of voxels \( C_1 \) and \( C_2 \), Dice's coefficient is defined as twice the number of overlapping voxels divided by the sum of the cardinalities of the two sets:
\[ D(C_1,C_2) = \frac{2 \times |C_1 \cap C_2|}{|C_1| + |C_2|}. \]
The Dice coefficient ranges from 0 to 1, where 0 indicates no overlap and 1 implies identical sets.
In our numerical studies, we perform this reproducibility assessment for tasks and resting states in both the ABCD and HCP studies. For each state, with post-preprocessing \( n \) subjects, we randomly select \( 75\% \) of the subjects 20 times to form a random sample set. For each sample set and each \( K = 200, 300, 400 \), the SBP algorithm is applied with the previously tuned optimal \( \lambda^* \) for \( K \), as detailed in Section \ref{sec:hyper}. We denote the resulting parcellation sets as \( \{\mathcal{C}^q, q= 1,2,\ldots,20\} \), where each \( \mathcal{C}^q = \{C_1^q,\ldots,C_K^q\} \) represents the partition results among $p$ voxels for the $q$ sample set. 
To align different parcellations to assess reproducibility, we then locate the most similar node from the remaining parcellations. For instance, for \( C_k^1 \), the \( k \)-th cluster from the first sample, we compute Dice's coefficient with all other nodes from each of \( \{\mathcal{C}^2,\ldots,\mathcal{C}^{20}\} \). The one with the highest Dice's coefficient is identified as the closest match, and the coefficients are recorded as \( D(k,1,q) \). The reproducibility for the \( k \)-th node from the current parcellation is then computed as the average of these 19 Dice's coefficients, denoted \( \bar{D}(k,1) = \frac{1}{19}\sum_{q=2}^{20} D(k,1,q) \). Ultimately, we summarize the average reproducibility score for each node, weighted by the proportion of the node's size to the total number of its voxels.

\section{Results}\label{sec:results}
\subsection{Simulated data}
We first demonstrate the proposed methods against standard spectral clustering using a simulated binary graph. This graph comprises 100 voxels arranged in a \(10 \times 10\) lattice with the corresponding adjacency matrix \( A \in \mathbb{R}^{100 \times 100} \). Based on the lattice, we have \( A_{ij} = 1 \) if voxel \( i \) and \( j \) are connected, and \( 0 \) otherwise. Simultaneously, we also define a separation preference matrix \( R \in \mathbb{R}^{100 \times 100} \) as follows--we set $r_{ij}$ to be a large value when $i\in (1, 2, \dots, 10)$ and $j\in (11, 12, \dots, 20)$; and we also set a large $r_{i,i+1}$ when \( i \) takes from the set $(8, 18, 28, \dots, 98)$. The remaining elements of \( R \) are set to \( 0 \). These non-zero entries in matrix \( R \) divide the voxels into distinct groups while maintaining spatial contiguity. Ideally, we want to prevent merging the voxels with a nonzero $r_{ij}$ into the same parcel to retain the highly predictive connectivity signals.  

\begin{figure}[H]
    \centering
\begin{subfigure}{.33\textwidth}
  \centering
  \includegraphics[width=1\linewidth]{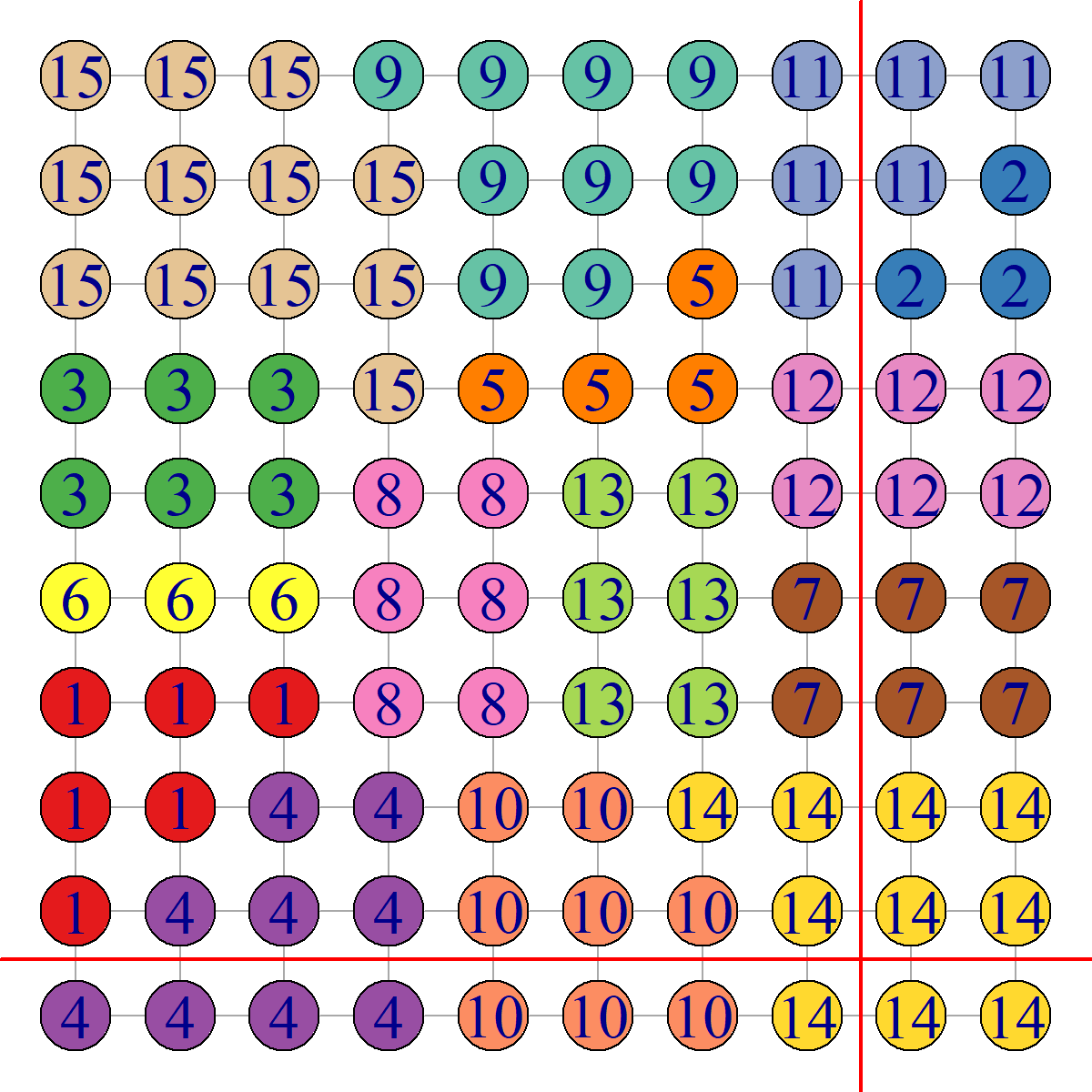}
  \caption{$\lambda = 0$}
  \label{fig:sim-lam0}
\end{subfigure}%
\begin{subfigure}{.33\textwidth}
  \centering
  \includegraphics[width=1\linewidth]{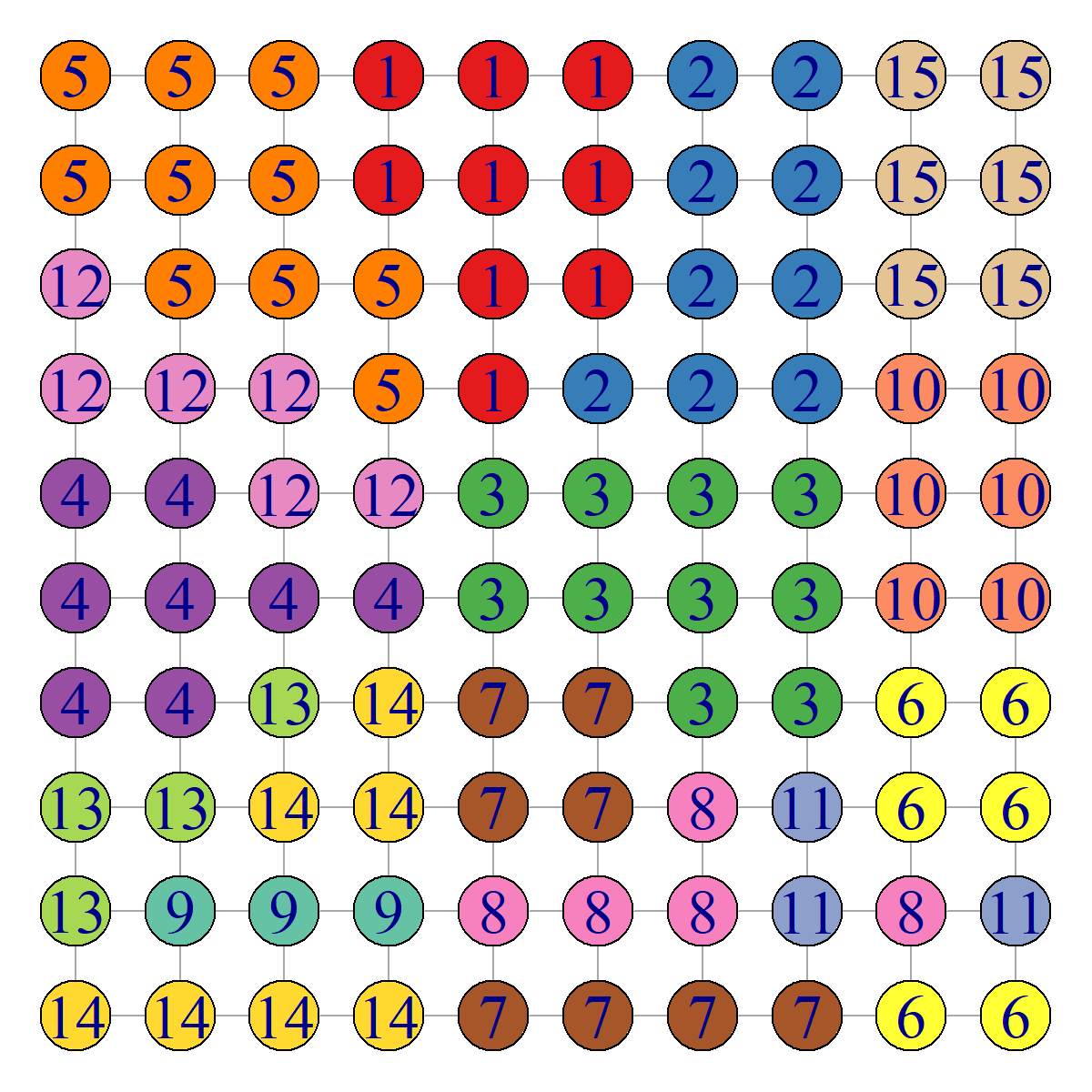}
  \caption{$\lambda = 5$}
  \label{fig:sim-lam5}
\end{subfigure}%
\begin{subfigure}{.33\textwidth}
  \centering
  \includegraphics[width=1\linewidth]{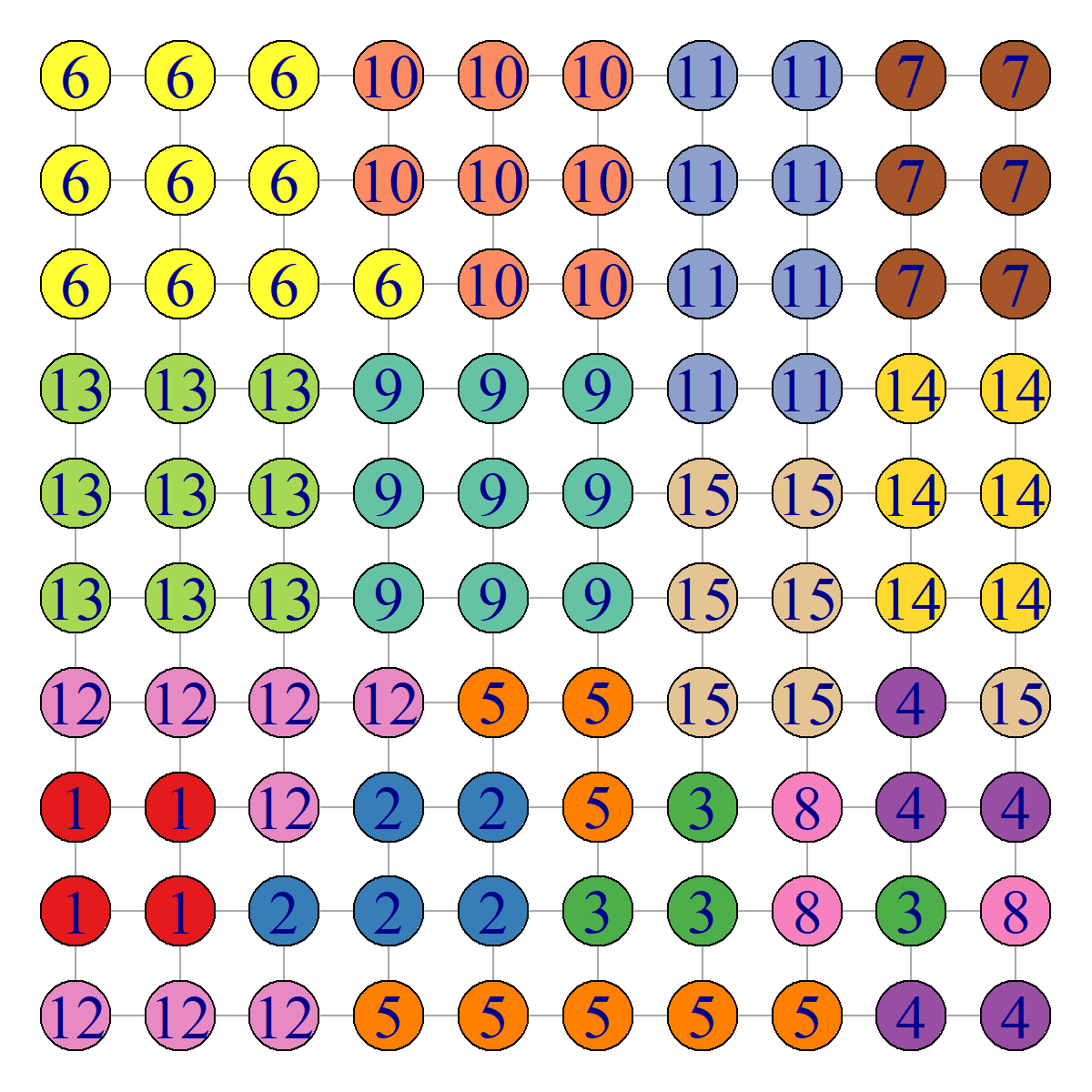}
  \caption{$\lambda = 10$}
  \label{fig:sim-lam10}
\end{subfigure}%
\caption{Demonstration of SBP with different $\lambda$ values under simulated data. The voxels cluster into the same parcel are assigned the same label and visualized by the same color. The red line segments indicate the high $R_{ij}$ pairs that we ought to prevent from merging. 
}
\label{fig:clustering-example}
\end{figure}

We apply the proposed SBP algorithm on the simulated data with \( \lambda \in \{0, 5, 10\} \),  and the resulting parcellations are displayed in Figure \ref{fig:clustering-example}. In the figure, voxels grouped in the same parcel are labeled with the same number and different clusters are distinguished by colors. When \( \lambda = 0 \), the proposed model reduces to standard spectral clustering marginally on \( A \), focusing solely on spatial proximity's spectral information and neglecting the predictive influence of the network on the outcome.  As \( \lambda \) increases, SBP method effectively segregates to maintain the predictive relevance of the constructed edges, while still achieving spatially coherent clustering. 
When \( \lambda = 5 \), clusters 6, 7, 8, 11 and 14 contain voxels that are not directly connected. Such a pattern is also observed for clusters 3, 4, 5, 8, 12 and 15 when \( \lambda = 10 \). 
These results are anticipated given that the penalty parameter associated with the separation-enforced term introduces a trade-off between the marginal pattern within the adjacency matrix and the predictive power linked with the targeted outcome. Compared to standard spectral clustering, SBP demonstrates a more nuanced approach, balancing spatial patterns with prediction-driven network separation.

\subsection{Predictive Performance}
We assess the connectome-based prediction for the cognitive scores based on the proposed SBP as compared to the existing parcellation schemes. We implement the proposed SBP with $K \in \{100,200,300,400,500\}$. As comparisons, we also consider existing atlases  including the AAL \citep{rollsAutomatedAnatomicalLabelling2020}, Shen268 atlas \citep{shenGroupwiseWholebrainParcellation2013}, Shen368 atlas and Schaefer atlas with $400$ parcels \citep{schaeferLocalGlobalParcellationHuman2018}. 
Each approach is evaluated under three states for the  ABCD study (Rest, MID, nBack) and HCP study (Rest, Language, Emotion) to ensure reliability under different conditions. 
Given there is no tuning parameter involved in the existing atlases, when evaluating their connectivity-based prediction, we directly construct the region-wise connectivity matrix based on the specific atlas and apply it to each of the testing sets as described in Section \ref{sec:hyper}. This ensures a fair comparison between our SBP and the existing ones with a consistent downstream prediction under CPM.
\begin{table}[!htb]
\centering
\caption{Predictive performance as summarized by the out-of-sample $R^2$ under SBP with different parcellation sizes and existing atlases for ABCD and HCP studies. } 
\label{tab:r2-abcd-hcp}
\begin{tblr}{
vspan=even,
  cells = {c},
  cell{1}{1} = {r=2}{r},
  cell{1}{2} = {r=2}{},
  cell{1}{3} = {c=3}{},
  cell{1}{6} = {c=3}{},
  cell{3}{1} = {c=2}{},
  cell{4}{1} = {r},
  cell{5}{1} = {r},
  cell{6}{1} = {r},
  cell{7}{1} = {r},
  cell{8}{1} = {r},
  cell{9}{1} = {c=2}{},
  cell{10}{1} = {r},
  cell{11}{1} = {r},
  cell{12}{1} = {r},
  cell{13}{1} = {r},
  vline{3} = {1-2}{},
  vline{4} = {1}{},
  vline{6} = {1-2}{},
  vline{3,6} = {4-8,10-13}{},
  hline{1,14} = {-}{0.2em},
  hline{2} = {3-8}{},
  hline{3-4,9-10} = {-}{}
}
Atlas                    & \# Regions ($K$) & ABCD           &                &                & HCP            &                &                \\
                         &            & Rest           & MID            & nBack          & Rest           & Language       & Emotion        \\
\textit{Proposed method} &            &                &                &                &                &                &                \\
SBP                      & 100        & 0.064          & 0.086          & 0.078          & 0.045          & 0.117          & 0.109          \\
SBP                      & 200        & 0.075          & 0.105          & 0.094          & 0.058          & 0.124          & 0.119          \\
SBP                      & 300        & 0.088          & 0.106          & 0.103          & 0.062          & 0.132          & 0.131          \\
SBP                      & 400        & 0.092          & 0.112          & \textbf{0.107} & \textbf{0.074} & \textbf{0.137} & \textbf{0.133} \\
SBP                      & 500        & \textbf{0.095} & \textbf{0.114} & 0.103          & 0.071          & 0.135          & 0.129          \\
\textit{Existing atlases}  &            &                &                &                &                &                &                \\
AAL3                     & 170        & 0.004          & 0.055          & 0.045          & 0.005          & 0.072          & 0.066          \\
Shen268                  & 268        & 0.008          & 0.072          & 0.066          & 0.012          & 0.106          & 0.093          \\
Shen368                  & 368        & 0.004          & 0.082          & 0.057          & 0.037          & 0.115          & 0.106          \\
Schaefer                 & 400        & 0.005          & 0.077          & 0.059          & 0.029          & 0.112          & 0.107          
\end{tblr}
\end{table}
As a result, the out-of-sample R-squares are presented in Table \ref{tab:r2-abcd-hcp}; and based on the results, we conclude that the proposed SBP dramatically improves the predictive performance compared to the existing atlases across a range of region numbers.  Notably, SBP shows the most marked improvement in predictive accuracy during resting conditions in both studies. For instance, in the ABCD study, the out-of-sample R-square increases by more than 170 times with the application of SBP compared with the existing ones.  Moreover, the SBP consistently outperforms existing atlases across different task conditions, demonstrating its robust predictive capability through supervised connectivity biomarkers.
An interesting observation from our comparison between the two studies is that SBP leads to a more pronounced prediction improvement in the ABCD study than in the HCP study. This discrepancy could be attributed to the children participants in the ABCD study. The fMRI scans of children often present more noise and exhibit stronger sample heterogeneity \citep{ota2014comparison}. These factors contribute to a more challenging analytical scenario, yet SBP successfully refines connectivity-based predictions with more informative functional parcellations that existing atlases struggle to achieve.

Finally, we observe that larger values of \( K \) (400 and 500) generally yield better predictive performance compared to smaller ones. This suggests that when the network is too small, significant connections might be merged into single groups, potentially missing individual contributions. On the other hand, a network that is too large could lose its benefit by merging voxels into regions in terms of noise reduction and enhancement on interpretation. Additionally, both with our SBP method and other approaches, task-based fMRI shows higher predictive accuracy than resting-state data. This is in line with various previous studies \citep{greeneTaskinducedBrainState2018, chen2022shared} with task conditions showing a stronger connectivity-based predictive power. 

\subsection{Supervised brain nodes} 
We then visualize our generated supervised brain nodes in Table \ref{tab:abcd-hcp-parc-plots} with both $K=200$ (SBP(200)) and 400 (SBP(400)) under ABCD and HCP studies; and each node in the parcellation is color-coded. From left to right, the axial, sagittal, and coronal slices are visualized using Bioimage Suite (\url{https://www.nitrc.org/projects/bioimagesuite/}).
\input{results/abcd-hcp-parcellation-plots}
As a direct consequence of assigning voxels based on their predictive power instead of solely on spatial proximity, the SBP-induced nodes exhibit higher variations across the whole brain. Unlike the AAL or Shen268 atlases, which are limited by fixed node boundaries, the SBP method dynamically generates atlases that are behavior-informed, state-specific, and trait-specific. Meanwhile, despite the different numbers of nodes, the generated parcellation schemes still demonstrate a certain degree of coherence. For instance, under the resting state ABCD study, both SBP(200) and SBP(400) display large nodes in the inferior temporal gyrus as shown in the axial view. Across different states, larger nodes are noticeable in the temporal lobe, caudate nuclei, and hippocampus. Additionally, being a whole-brain parcellation method, the SBP ensures a degree of spatial contiguity through the adjacency matrix, as shown in the obtained brain nodes. Under different states, with distinct predictive roles that functional signals involved, our constructed parcellations are also thoroughly tailored.  


\subsection{Functional network at large-scale network level}\label{sec:fc-network}
Having established that the proposed SBP consistently enhances predictive accuracy for new samples, we further explore the reasons behind this improved performance and the established functional signals linked to the targeted outcome.
We consider the parcellations generated from SBP(400) as an illustrative example, given its consistently superior performance across ABCD and HCP studies and the comparable number of nodes with existing atlases. For each state and each study, we focus on the top \(20\%\) functional connections identified by the last step CPM, and map them to the ten canonical neural networks \citep{yeo2011organization} as shown in 
figure \ref{fig:abcd-hcp-cpm-plots}. 
In each subfigure of our analysis, we present heatmaps that show the number of these functional connections, both within and between the canonical neural networks. We separately summarize the heatmaps by distinguishing between connections that are positively and negatively associated with the outcome. 
\input{results/abcd-hcp-cpm}

The observed patterns in our analysis indicate that our parcellation schemes are effective at identifying both inter-network connections and intra-network ones, and the informative network neuromarkers vary across different states and study populations. In the ABCD study, the motor and default mode networks heavily contribute to both inter- and intra-network connectivity neuromarkers in positive networks across states; and the cerebellum network also plays a role on top of aforementioned networks in negative networks. In the HCP study, the medial frontal network is highlighted to offer inter- and intra-network connections under resting-state, while the fronto parietal network contributes substantially under task conditions along with the cerebellum. Of note, it is anticipated to observe different network signals between ABCD and HCP studies with distinct study cohorts. The functional involvement of the brain with behavior varies between preadolescents (ABCD study) and young adults (HCP study), reflecting developmental changes and maturity levels in these distinct age groups

Furthermore, we examine the canonical neural network distribution of the top 5 largest nodes identified by SBP(400) under both studies, presented in Figure \ref{fig:abcd-hcp-top5roi}.  The results indicate a diverse functional network composition within these top nodes, and the architectural patterns also vary across different cognitive conditions and studies. For example, in the resting state of ABCD study, the second-largest node contains \(13.56\%\) of its voxels from the cerebellum network (highlighted in yellow), which represents the largest percentage of cerebellum among the top five nodes. During the MID and nBack tasks, node 5 and node 3 comprise \(19.06\%\) and \(21.13\%\) of voxels within the cerebellum network, respectively. We also observe that the proportion of Visual II (indicated in purple) is significantly lower in general under the nBack task compared to the resting state. In the HCP study, node 3 in the emotion task comprises \(18.65\%\) of its voxels within the basal ganglia (marked in teal) network, whereas the largest node from the Language task contains only \(11.86\%\). Such diversity in patterns and network compositions highlights the supervised nature of our brain parcellation approach. Unlike existing atlas methodologies that often strive for intra-node functional homogeneity, our method acknowledges and integrates the functional diversity within each node to enhance the predictive power for behavior outcomes.
\input{results/abcd-hcp-top5roi}

\subsection{Functional network anatomy}
To illustrate the informative network configurations constructed by SBP(400) in relation to the macroscale brain structure, we focus on the informative functional connections with the outcome under a correlation strength larger than 0.1  for each state and each study. These informative networks are displayed in 
Figure \ref{fig:abcd-hcp-ringplots}, where we separately present positive and negative correlated connections. 
In each subfigure, nodes are arranged in two semicircles, approximating the brain's anatomy from anterior (top of the circle, 12 o'clock position) to posterior (bottom of the circle, 6 o'clock position), color-coded according to cortical lobes. The nodes (inner circle) are anatomically grouped into lobes (outer circle) split into left and right hemispheres with each line representing an informative connection. 

Based on the result, similar to previously identified connectivity features (e.g. \cite{finn2015functional}), the behavior-related network configurations are generally complex and span across various brain macroscale areas. In each state and each study, we observe a large proportion of long-range, cross-hemisphere connections in both positive and negative networks, which demonstrates their dominant predictive role for general cognitive ability. Among those connections, many are between different lobes or occur within a single lobe but across hemispheres. For instance, the functional networks linking parietal and temporal lobes are particularly prominent in resting and nBack states for the ABCD study, and in the resting and language states for the HCP study. The connections between the parietal and motor are also extensively observed in most conditions under both studies. Furthermore, there are a higher number of informative connections in certain areas such as prefrontal, motor and parietal involved compared to the occipital, cerebellum and subcortical lobes, especially during tasks.

\input{results/abcd-hcp-ringplots}

\subsection{Reproducibility}
In this section, we evaluate the reproducibility of the supervised brain nodes generated by SBP under ABCD and HCP studies. As detailed in Section \ref{sec:reprod}, we calculate the averaged Dice's coefficient for each node across 20 random samples, and summarize the distribution of these coefficients separately under the left and right hemispheres to examine how they vary across different numbers of nodes, states and studies as shown in Figure \ref{fig:abcd-hcp-reproducibility}. In addition to Dice's coefficients, under each random sample, we also evaluate the robustness of the node size by calculating the number of voxels in each node, and summarizing the distributions of its averaged values across all the nodes for each state and study.

\input{results/abcd-hcp-reproducibility}

As shown in the left panels, the proposed parcellation process consistently yields comparable node sizes between two hemispheres under all the tasks and \( K \) values. 
This indicates balanced parcellation configurations between hemispheres with minimal variations across random samples. Regarding reproducibility, as shown in the right panel of the figure, the Dice's coefficients we obtained are moderate, which is a reasonable outcome considering the nature of the measurement. The calculation of Dice’s coefficient involves aligning parcellations from different random samples, a process that can be challenging due to the flexible nature of node definitions in our method. While this flexibility may lead to some degree of loss in reproducibility, the moderate levels of Dice's coefficients suggest an acceptable level of consistency in the parcellation outcomes across different studies and conditions.

\section{Discussion}\label{sec:discussion}
In this work, we introduce the SBP framework, a novel approach designed to meet the urgent demand for more informative functional brain parcellations, particularly in the context of predictive tasks. Our framework uniquely defines brain functional nodes and their networks, taking into account their relationship with relevant behavioral outcomes. Our analytical framework is built under a regularized spectral clustering algorithm, which facilitates graph-based community detection while integrating behavioral outcome associations through a separation preference matrix. Our extensive numerical analyses, using both simulated data and two landmark multi-state fMRI studies, consistently highlight the superior performance of SBP in enhancing connectome-based predictions. This advancement not only outperforms predictions based on existing brain atlases but also lays the groundwork for establishing behavior-refined and more informative functional brain nodes and networks. 

Our optimization function currently integrates both an unsupervised spectral learning of connectivity data and a separation procedure guided by the behavior outcome. The spectral learning component inherently promotes some spatial smoothness. However, with an emphasis on the behavior-guided component, a single node could separate into non-contiguous segments, without a guarantee of spatial continuity. While traditional brain atlases focus on spatially contiguous regions, such contiguity is not considered to be a strict requirement. Nevertheless, spatially contiguous functional nodes do offer enhanced interpretability.
In light of this, a potential improvement to our SBP framework could involve the integration of spatial information into the objective function via a spatial regularization term. Such an enhancement would not only preserve the method's predictive power but also improve its anatomical interpretability.


Another potential extension is on the construction of both the connectivity adjacency matrix \( A \) and the regularization matrix \( R \). Currently, both matrices rely on Pearson correlation for measuring voxel-wise similarity and edge-behavior association, respectively. While Pearson correlation offers straightforward numerical operation and interpretation, it primarily captures linear relationships, potentially overlooking complex higher-order and nonlinear correspondences. Therefore, to address this potential limitation, kernel-based or neural-network-based approaches could be developed as alternative options to provide more nuanced characterizations of those complex relationships. Furthermore, we currently focus on predicting the general cognitive score as a canonical behavior outcome to demonstrate the efficacy of our framework. Going forward, it would be beneficial to broaden our scope to include other behavioral traits and disease profiles to establish more tailored functional nodes and networks that are relevant to a wider array of behavioral and clinical contexts.


\section*{Ethics}
The secondary data analysis of ABCD and HCP data in this work was covered under IRB 2000020891.

\section*{Data availability}
This work adopted publicly available data from the ABCD and HCP studies. The ABCD data are available through the NIMH Data Archive (NDA). https://nda.nih.gov/, and the HCP data can be accessed via ConnectomeDB (https://db.humanconnectome.org). The code for the proposed SBP algorithm is available at https://github.com/wanwanx/SBP. 

\section*{Author Contributions}
W.X., T.C. and Y.Z. designed the study. W.X., T.L., and Y.Z. designed the algorithm. W.X., S.W., W.L., and C.T. processed and analyzed the data. All authors contributed to interpreting the results and writing the paper. 

\section*{Funding}
This work was partially supported by the National Institutes of Health under awards RF1AG081413, R01EB034720 and  RF1AG068191.

\section*{Declaration of Competing Interest}
The authors report no competing interests.

\bibliographystyle{apalike}
\bibliography{BrainParcellation.bib}

\end{document}

%% file: format.tex
\usepackage{amssymb}
\usepackage{amsthm}

\usepackage{lineno}

\usepackage[figuresright]{rotating}

\newcommand{\x}{\mathbf{x}}

\theoremstyle{plain}

\theoremstyle{definition}

\usepackage{amsmath}
\usepackage{amsthm}
\usepackage{amssymb}
\usepackage{setspace}
\usepackage{graphicx, color}
\usepackage{algorithm}
\usepackage{algpseudocode}
\usepackage{mathrsfs} 
\usepackage{lipsum}
\usepackage{booktabs}
\usepackage{multirow}
\usepackage{graphicx}

\usepackage{caption}
\usepackage{subcaption}

\usepackage[colorlinks=true,linkcolor=blue]{hyperref}%
\usepackage{array, blindtext}

\usepackage[dvipsnames]{xcolor}

\usepackage{enumitem}

\usepackage{tabularray}

%% file: results/abcd-hcp-parcellation-plots.tex
\begin{table}[!ht]
  \centering
 \begin{tabular}{  m{1.5cm} | c | c }
\toprule
 \rule{0pt}{20pt} & \textbf{SBP(200)} & \textbf{SBP(400)} \\
    \hline
\multicolumn{3}{c}{ \hspace{+1.5cm}\textit{ABCD}} \\ 
\hline
    \textbf{Rest} 
    &
    \begin{minipage}{.43\textwidth}
      \includegraphics[width=0.95\textwidth]{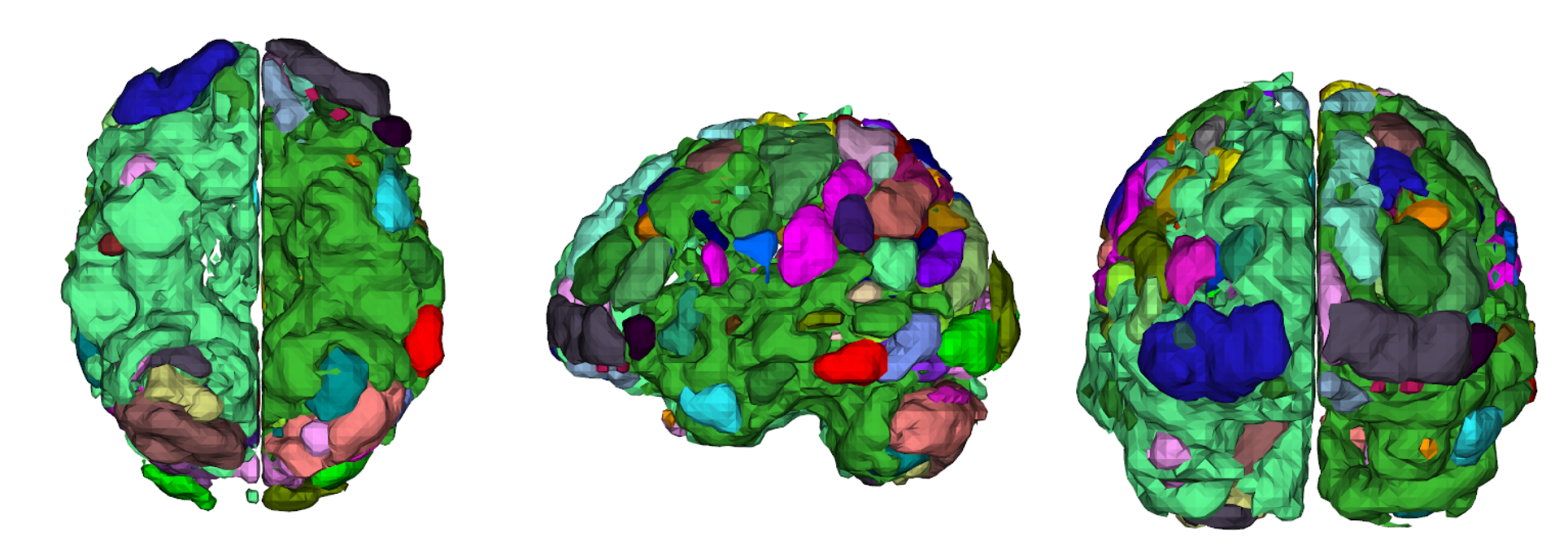}
    \end{minipage}
     &
    \begin{minipage}{.43\textwidth}
      \includegraphics[width=0.95\textwidth]{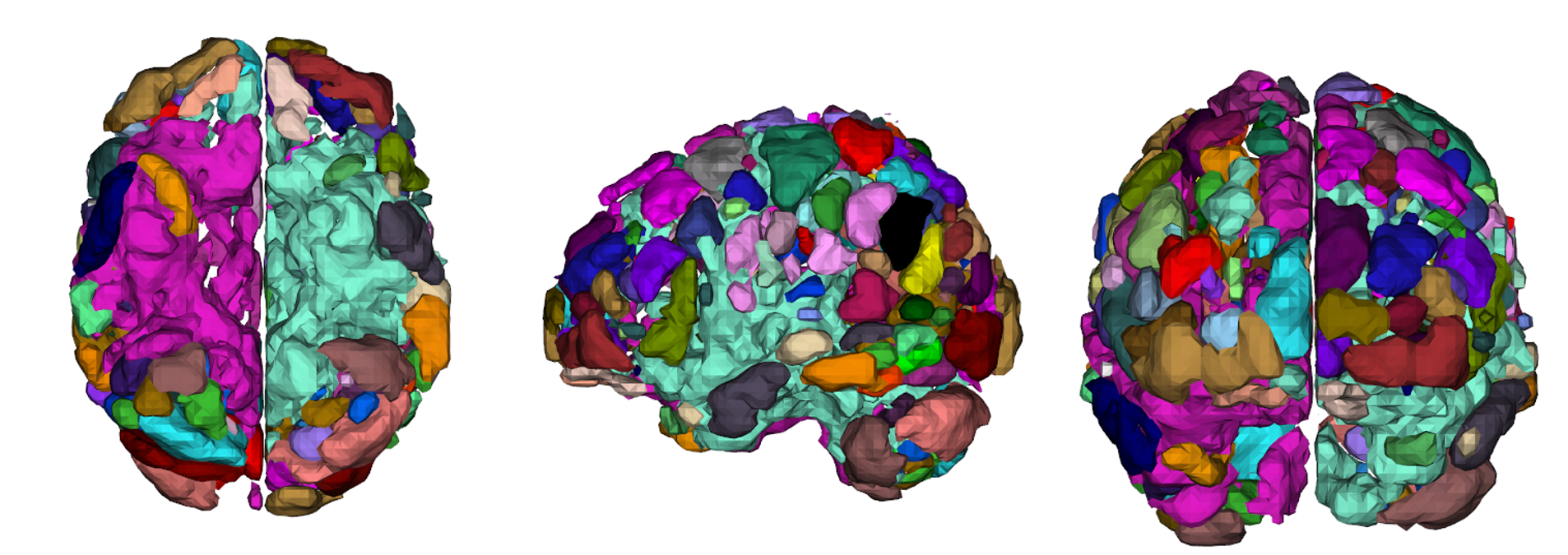}
    \end{minipage}
    \\ \hline
    \textbf{MID}
    &
    \begin{minipage}{.43\textwidth}
      \includegraphics[width=0.95\textwidth]{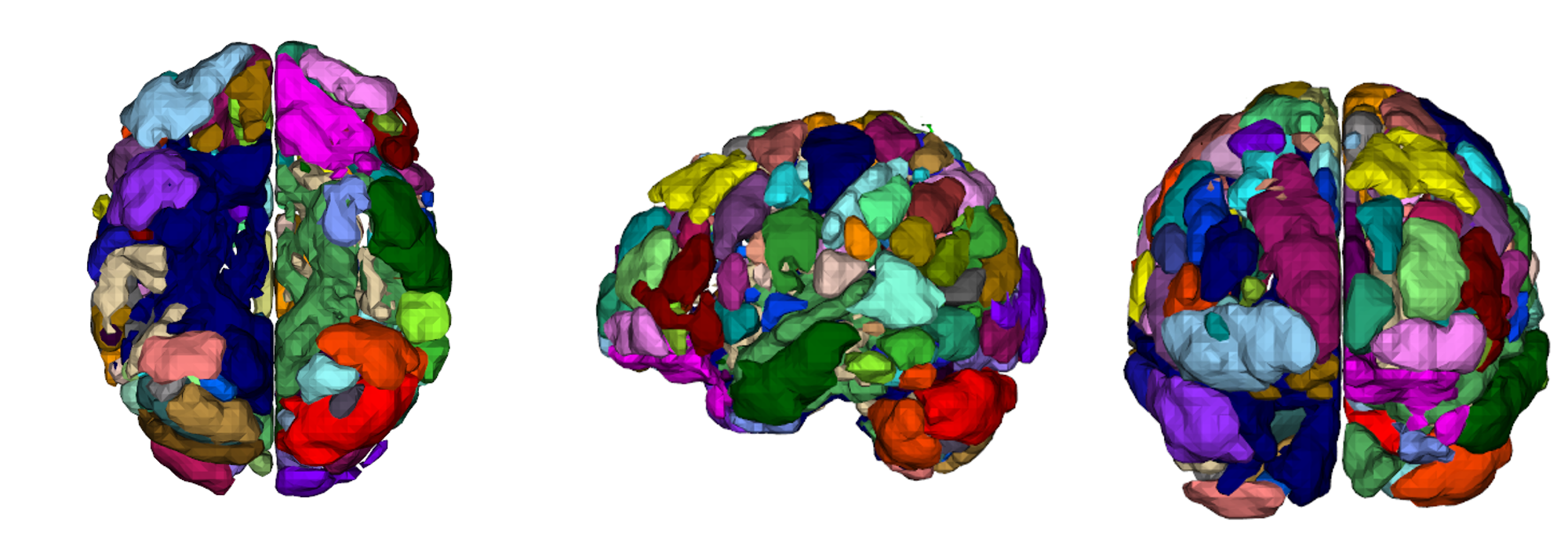}
    \end{minipage}
     &
    \begin{minipage}{.43\textwidth}
      \includegraphics[width=0.95\textwidth]{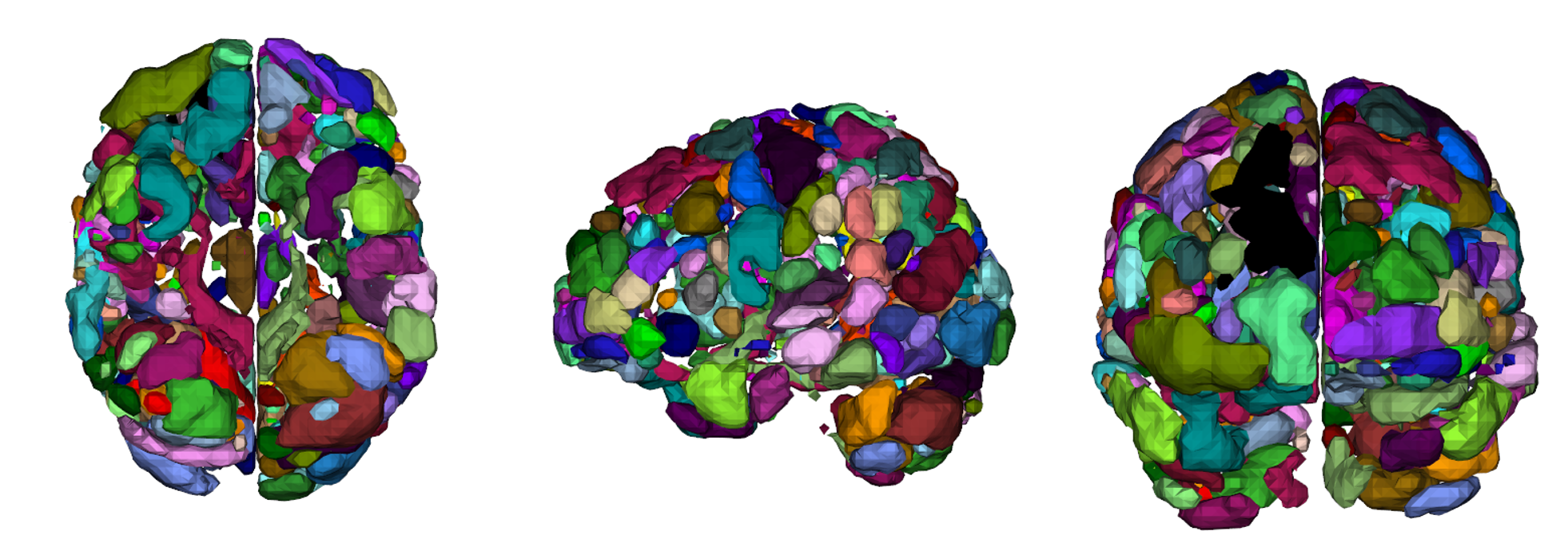}
    \end{minipage}
    \\ \hline
    \textbf{nBack} 
    &
    \begin{minipage}{.43\textwidth}
      \includegraphics[width=0.95\textwidth]{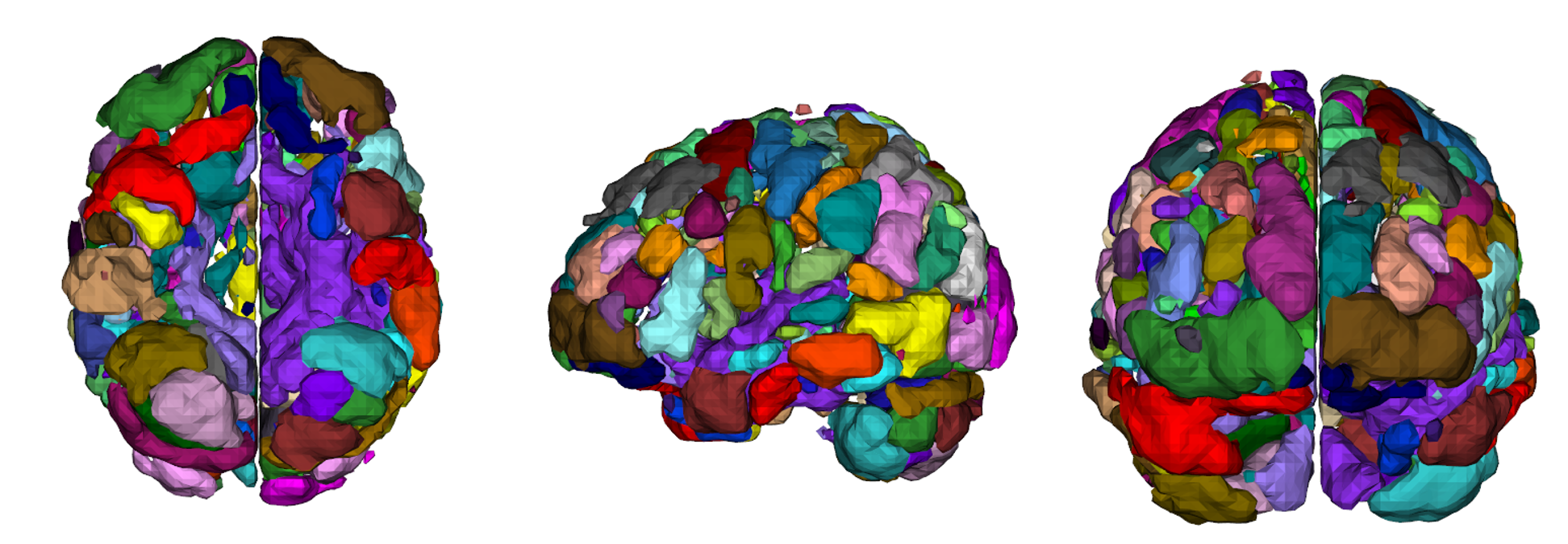}
    \end{minipage}
     &
    \begin{minipage}{.43\textwidth}
      \includegraphics[width=0.95\textwidth]{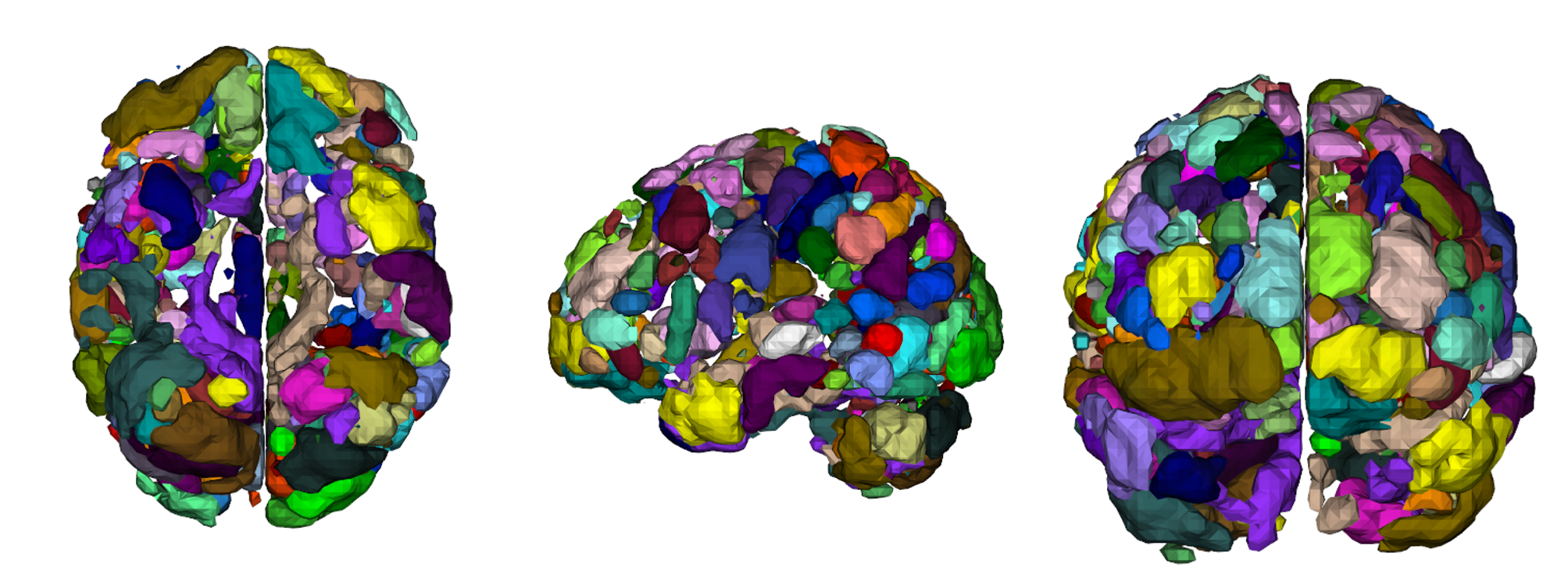}
    \end{minipage}
    \\ 
    \midrule
\multicolumn{3}{c}{ \hspace{+1.5cm}\textit{HCP}} \\ 
    \hline
    \textbf{Rest} 
    &
    \begin{minipage}{.43\textwidth}
      \includegraphics[width=0.32\textwidth]{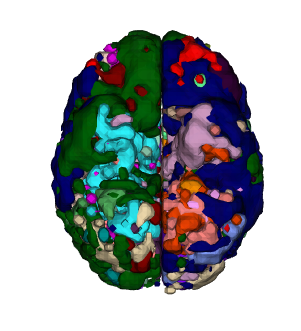}
      \includegraphics[width=0.32\textwidth]{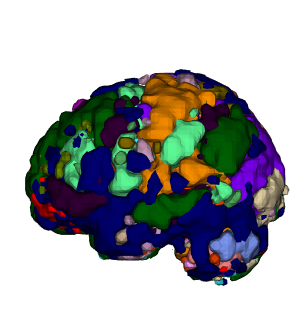}
      \includegraphics[width=0.32\textwidth]{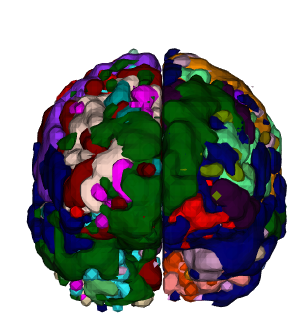}
    \end{minipage}
     &
    \begin{minipage}{.43\textwidth}
      \includegraphics[width=0.32\textwidth]{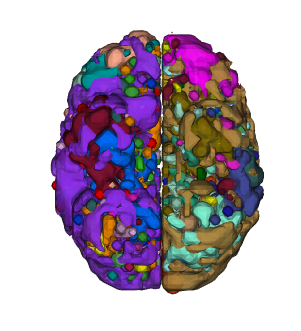}
      \includegraphics[width=0.32\textwidth]{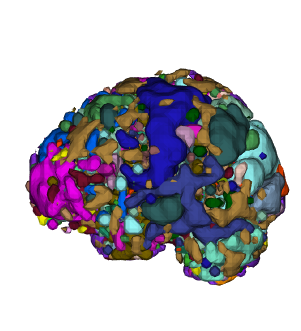}
      \includegraphics[width=0.32\textwidth]{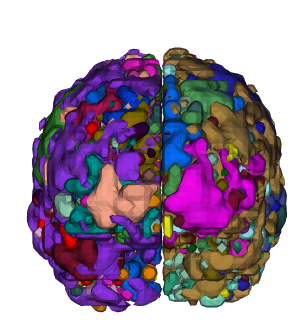}
    \end{minipage}
    \\ \hline
    \textbf{Language}
    &
    \begin{minipage}{.43\textwidth}
      \includegraphics[width=0.32\textwidth]{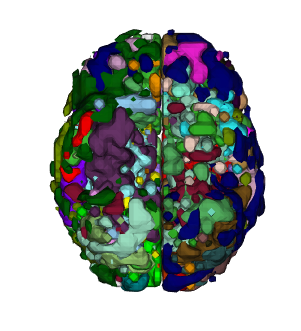}
      \includegraphics[width=0.32\textwidth]{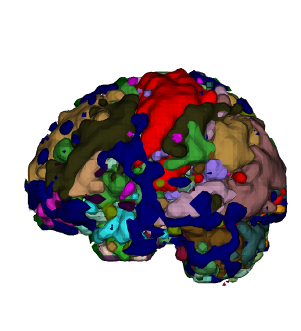}
      \includegraphics[width=0.32\textwidth]{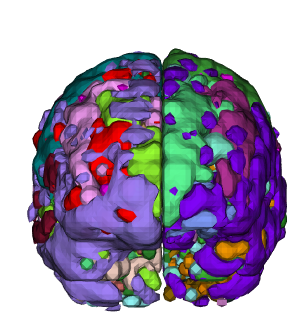}
    \end{minipage}
     &
    \begin{minipage}{.43\textwidth}
      \includegraphics[width=0.32\textwidth]{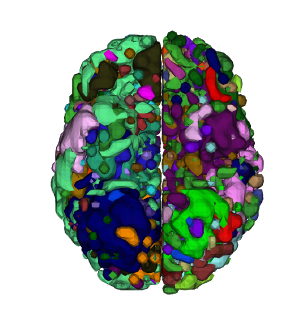}
      \includegraphics[width=0.32\textwidth]{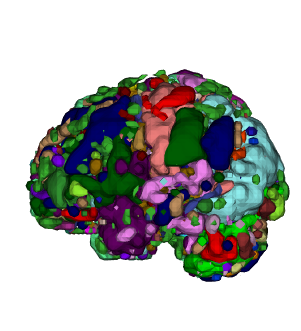}
      \includegraphics[width=0.32\textwidth]{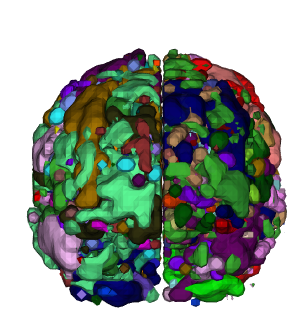}
    \end{minipage}
    \\ \hline
    \textbf{Emotion} 
    &
    \begin{minipage}{.43\textwidth}
      \includegraphics[width=0.32\textwidth]{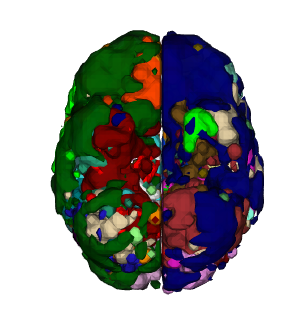}
      \includegraphics[width=0.32\textwidth]{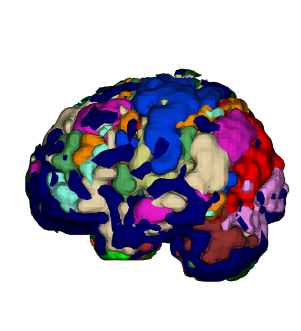}
      \includegraphics[width=0.32\textwidth]{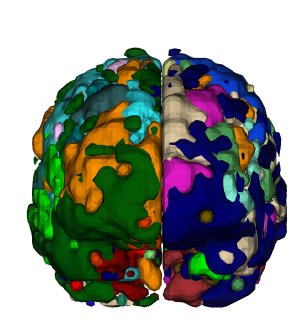}
    \end{minipage}
     &
    \begin{minipage}{.43\textwidth}
      \includegraphics[width=0.32\textwidth]{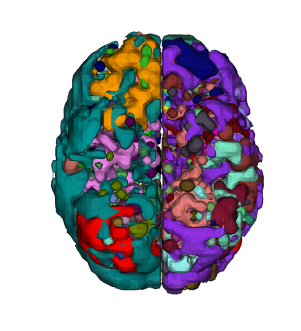}
      \includegraphics[width=0.32\textwidth]{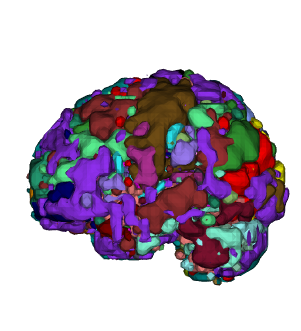}
      \includegraphics[width=0.32\textwidth]{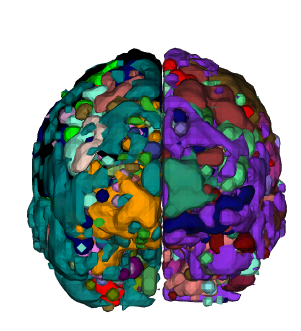}
    \end{minipage}
    \\ 
    \bottomrule
\end{tabular}
  \caption{Supervised brain nodes generated with SBP(200) and SBP(400) for intelligence scores under ABCD and HCP studies.}\label{tab:abcd-hcp-parc-plots}
\end{table}

%% file: results/abcd-hcp-cpm.tex
\begin{figure}[H]
    \centering
     \begin{subfigure}[t]{0.45\textwidth}
         \centering
         \includegraphics[width=0.48\textwidth]{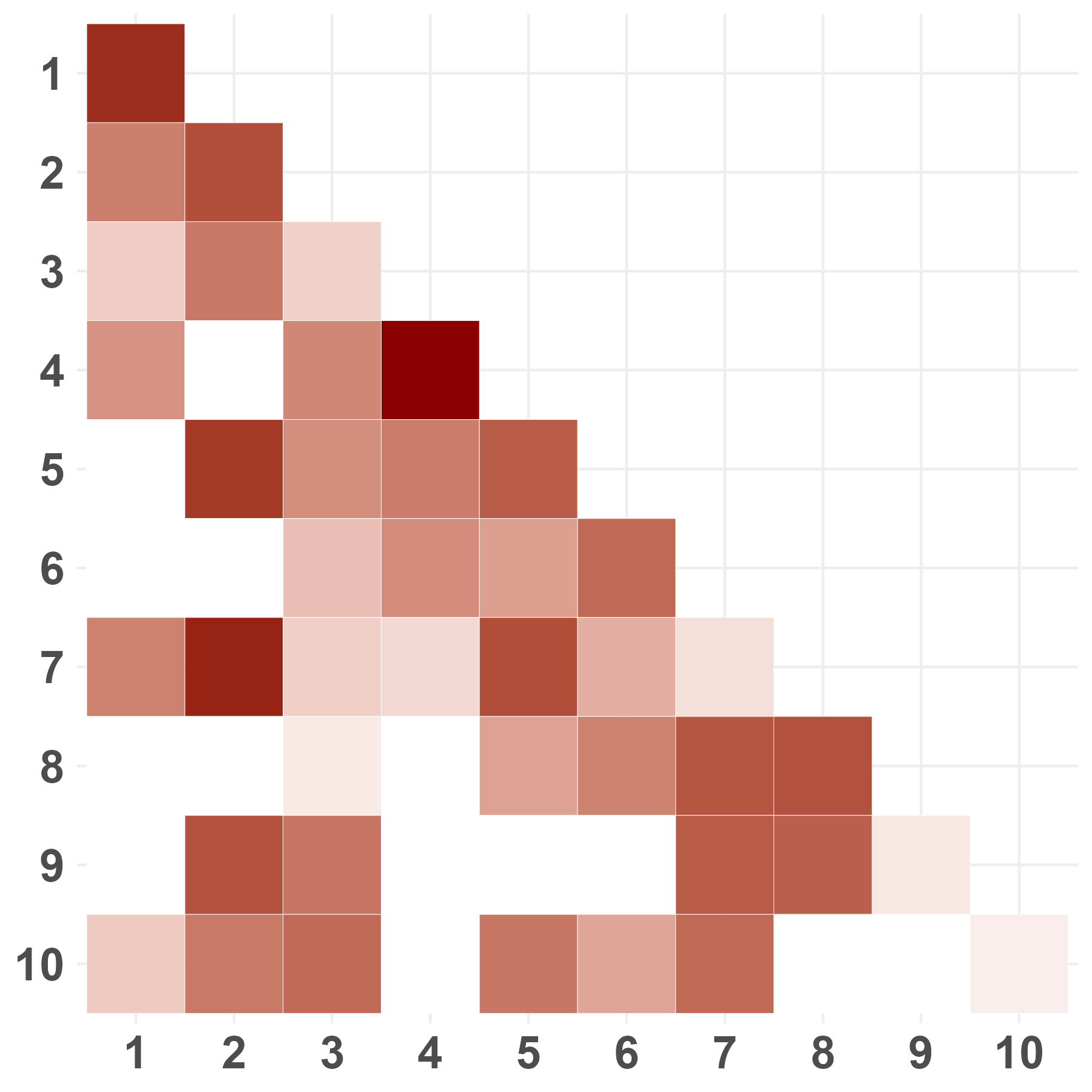}
         \includegraphics[width=0.48\textwidth]{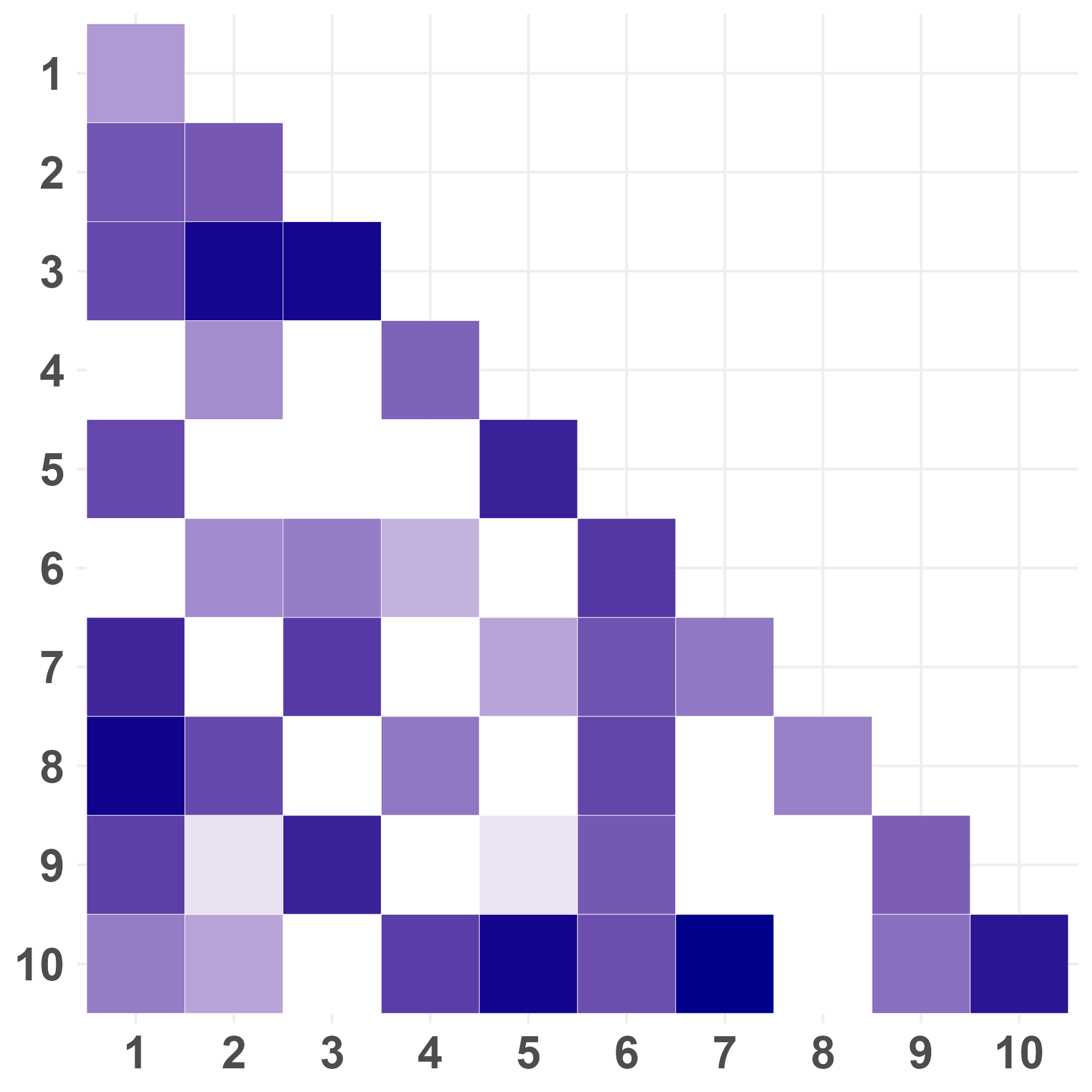}
         \caption{ABCD-Rest}
     \end{subfigure}
     \begin{subfigure}[t]{0.45\textwidth}
         \centering
         \includegraphics[width=0.48\textwidth]{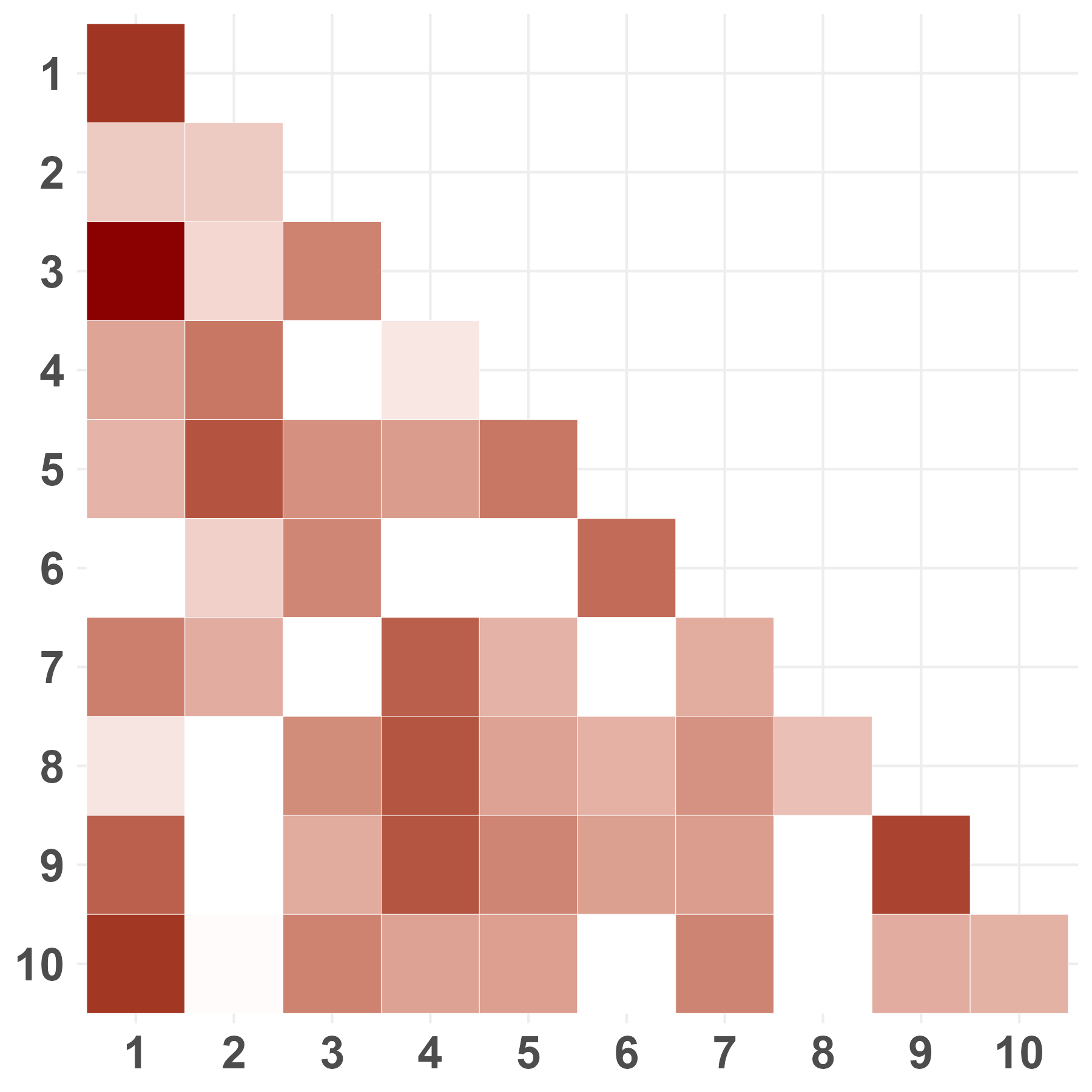}
         \includegraphics[width=0.48\textwidth]{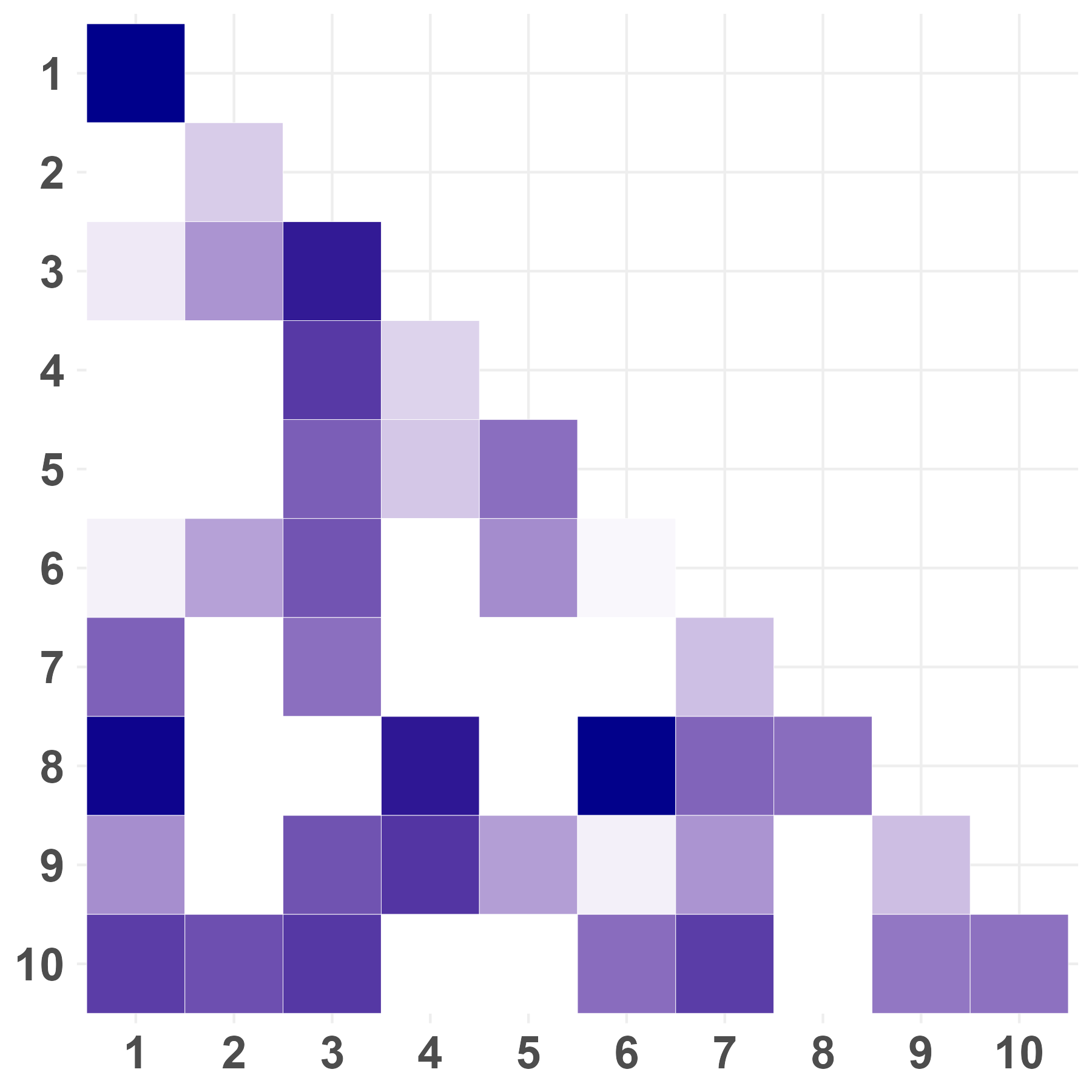}
         \caption{HCP-Rest}
     \end{subfigure}
     \vfill
     
     \begin{subfigure}[t]{0.45\textwidth}
         \centering
         \includegraphics[width=0.48\textwidth]{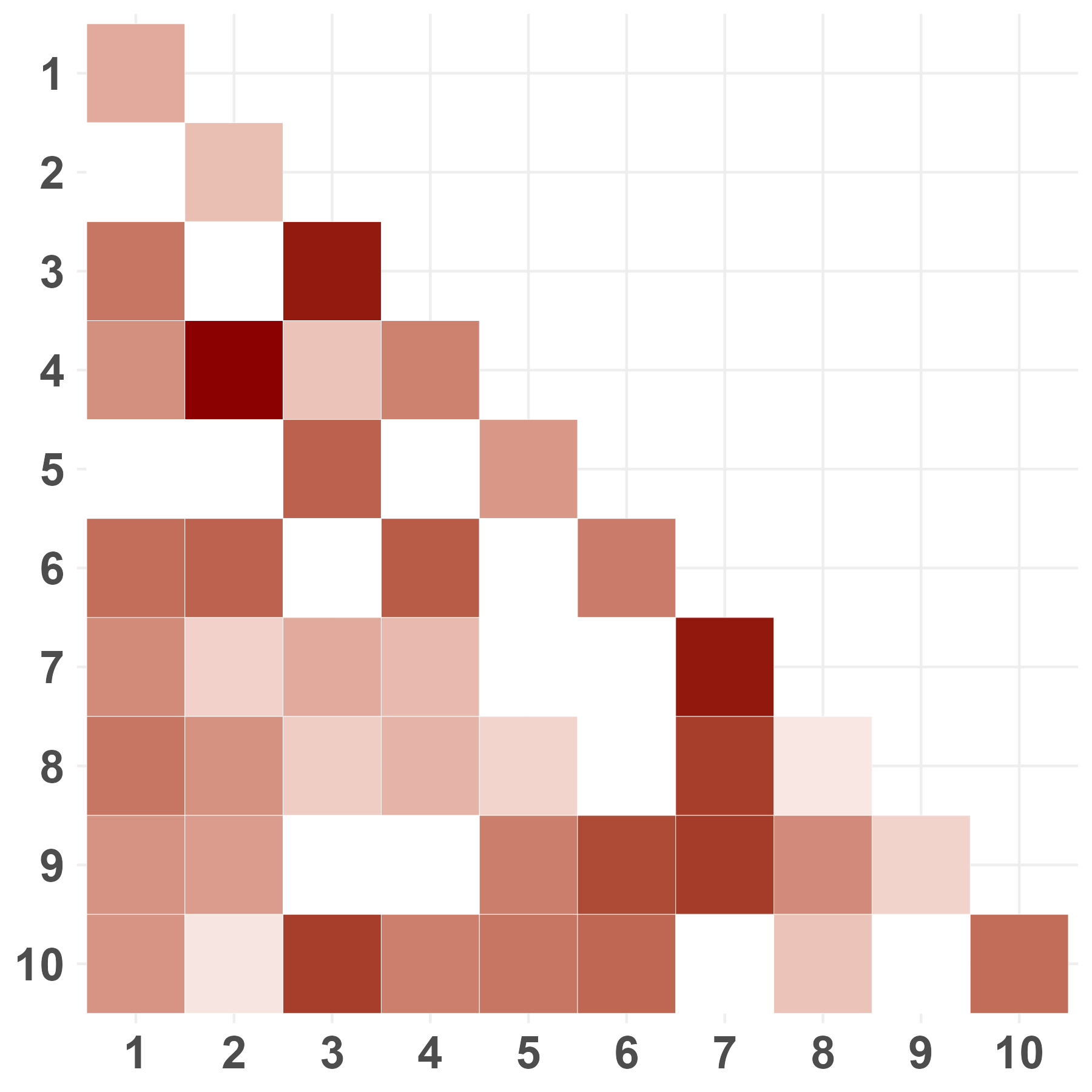}
         \includegraphics[width=0.48\textwidth]{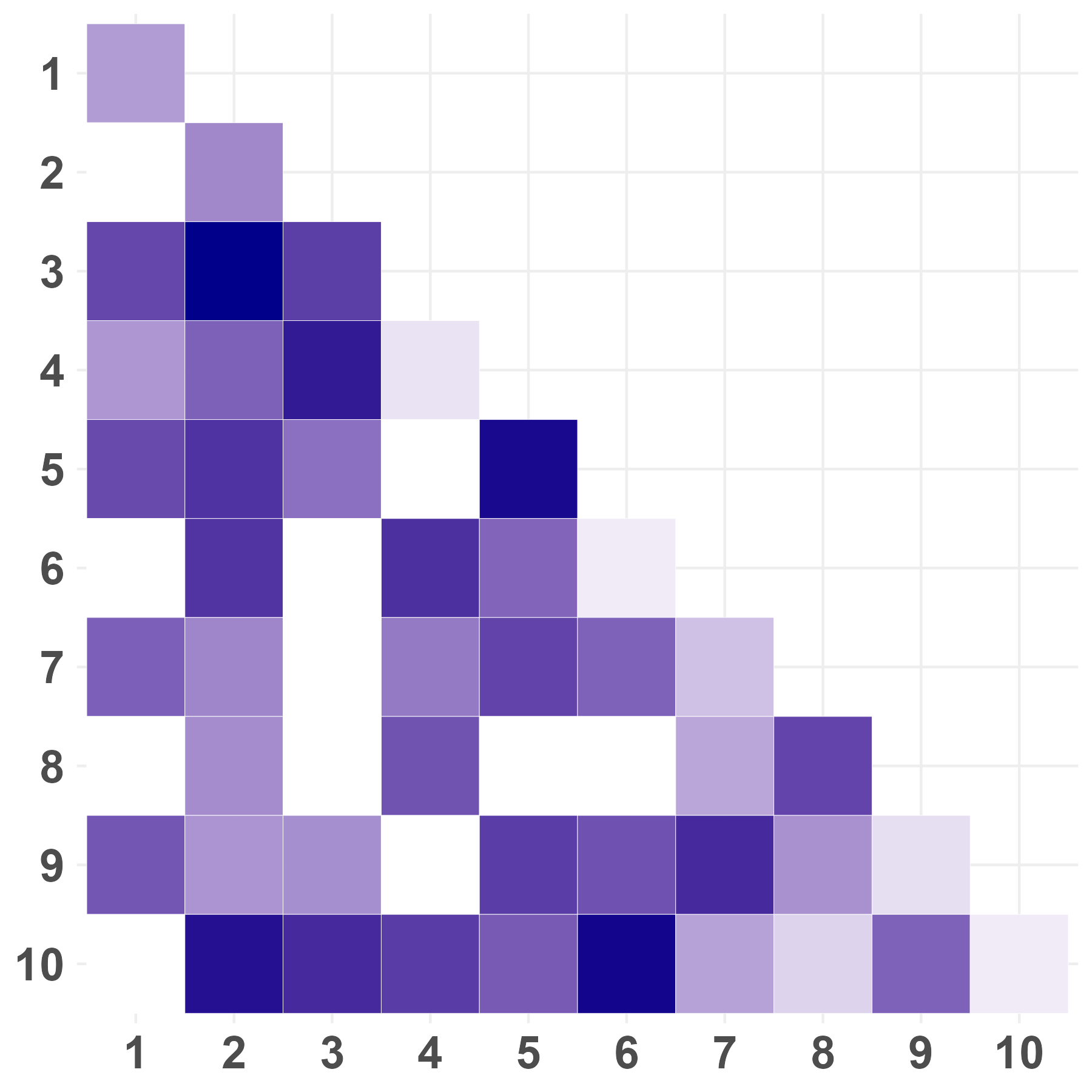}
         \caption{ABCD-MID}
     \end{subfigure}
     \begin{subfigure}[t]{0.45\textwidth}
         \centering
         \includegraphics[width=0.48\textwidth]{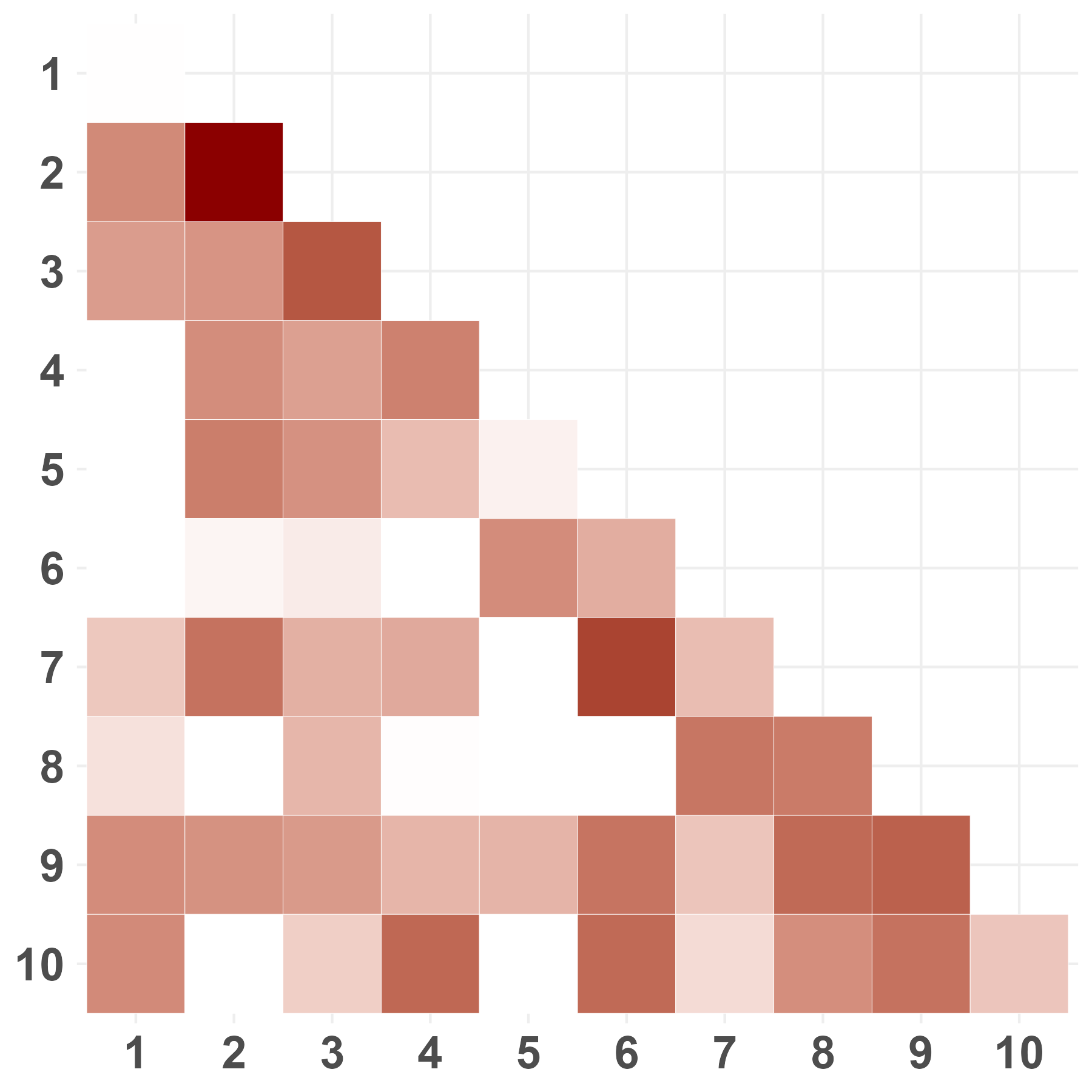}
         \includegraphics[width=0.48\textwidth]{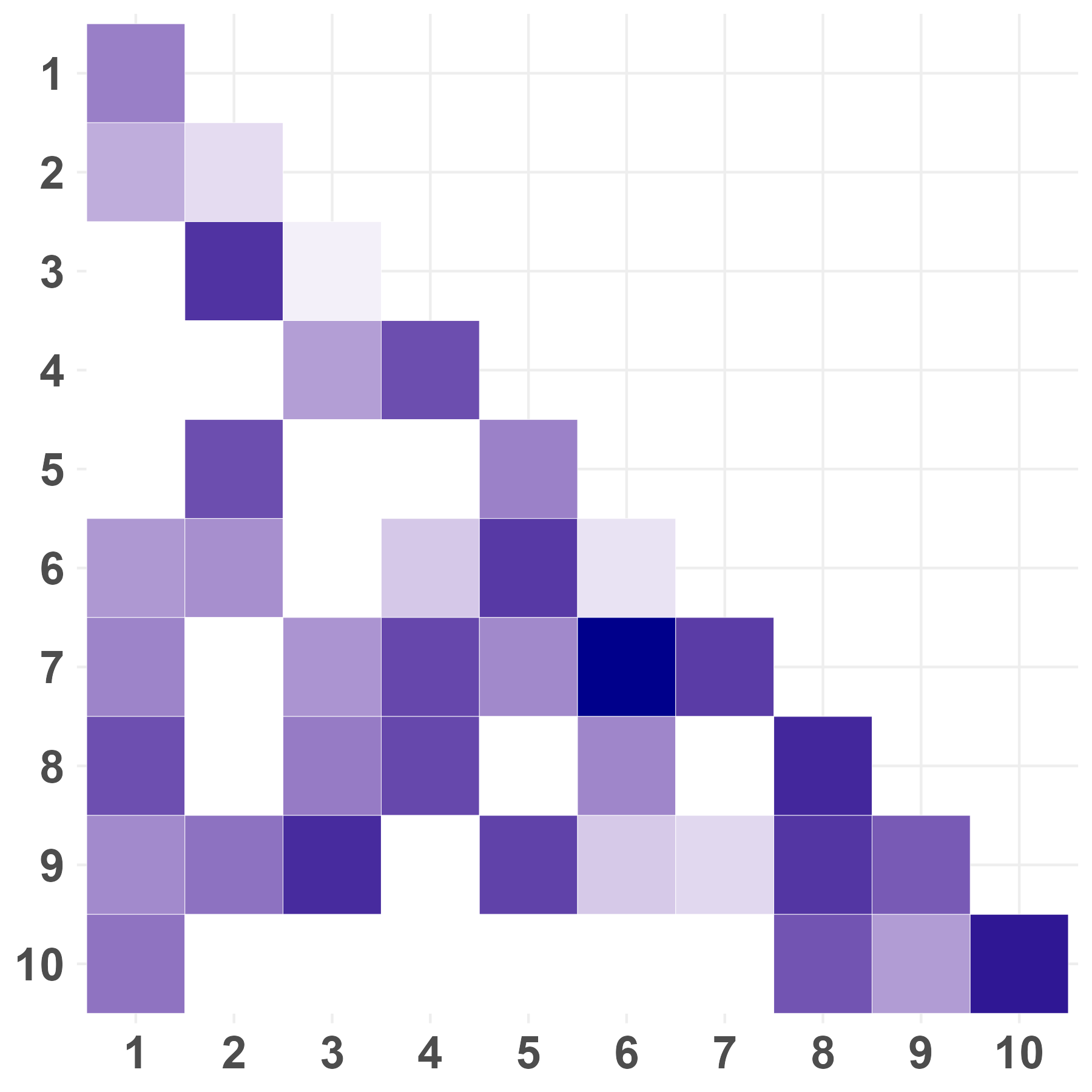}
         \caption{HCP-Language}
     \end{subfigure}
     \vfill

     \begin{subfigure}[t]{0.45\textwidth}
         \centering
         \includegraphics[width=0.48\textwidth]{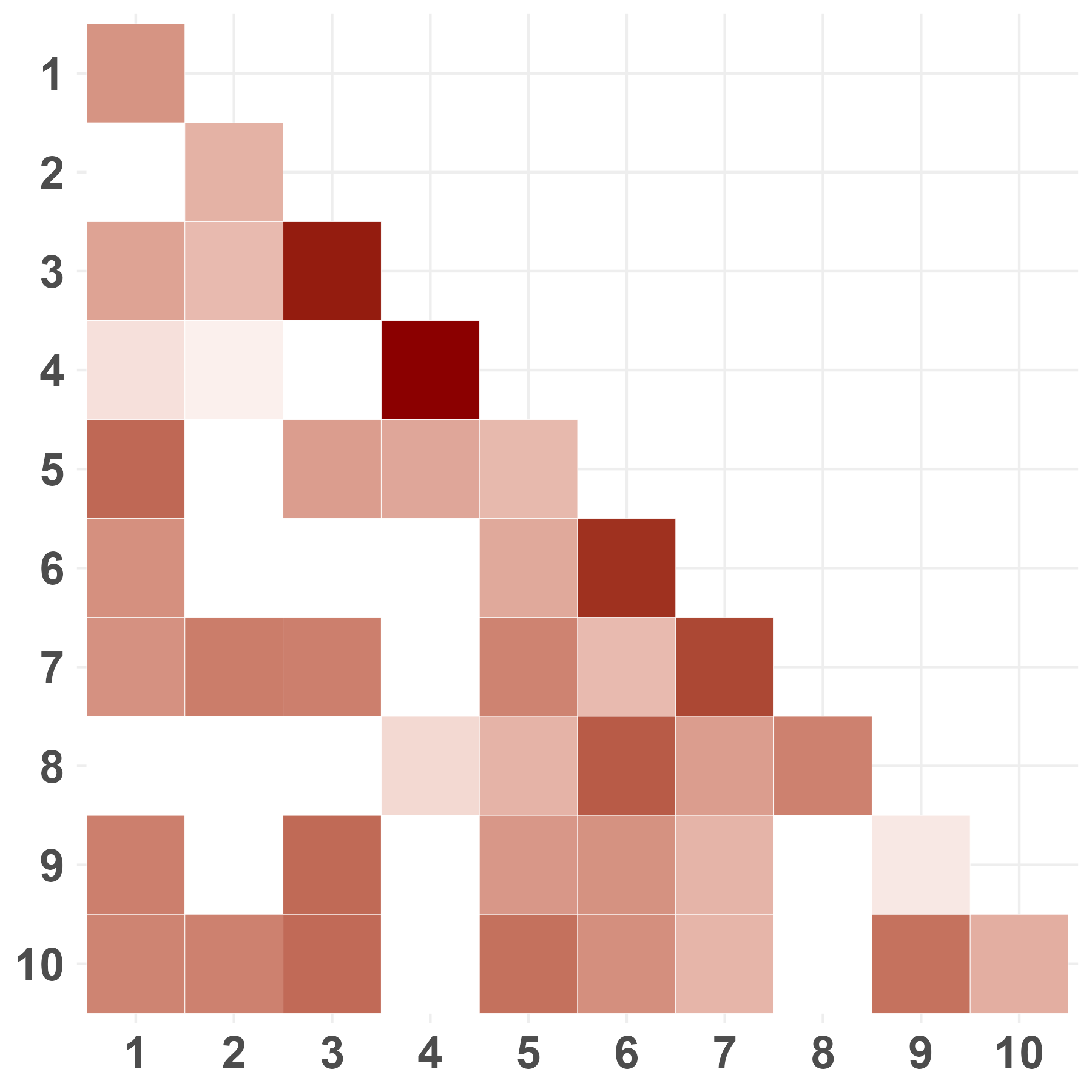}
         \includegraphics[width=0.48\textwidth]{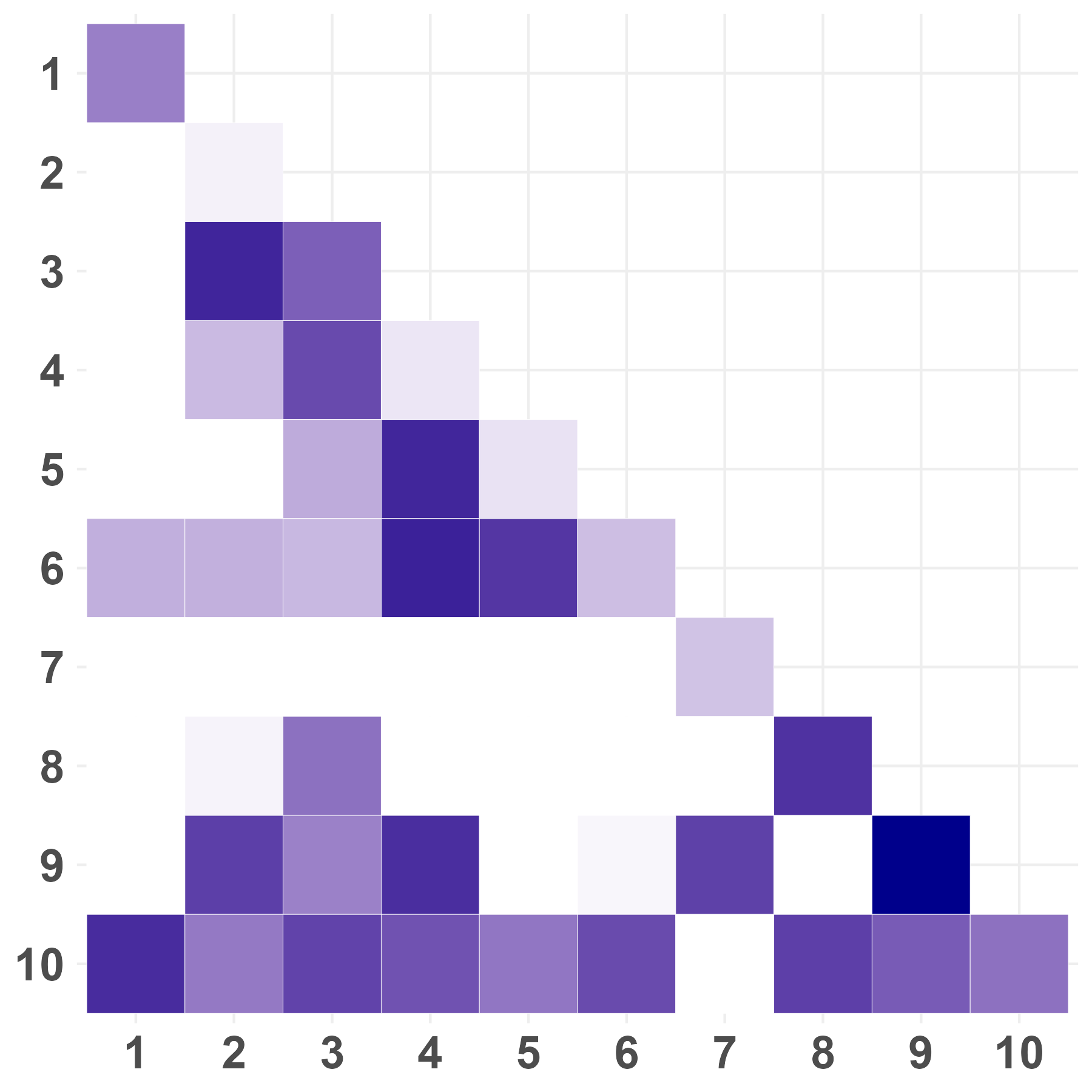}
         \caption{ABCD-nBack}
     \end{subfigure}
     \begin{subfigure}[t]{0.45\textwidth}
         \centering
         \includegraphics[width=0.48\textwidth]{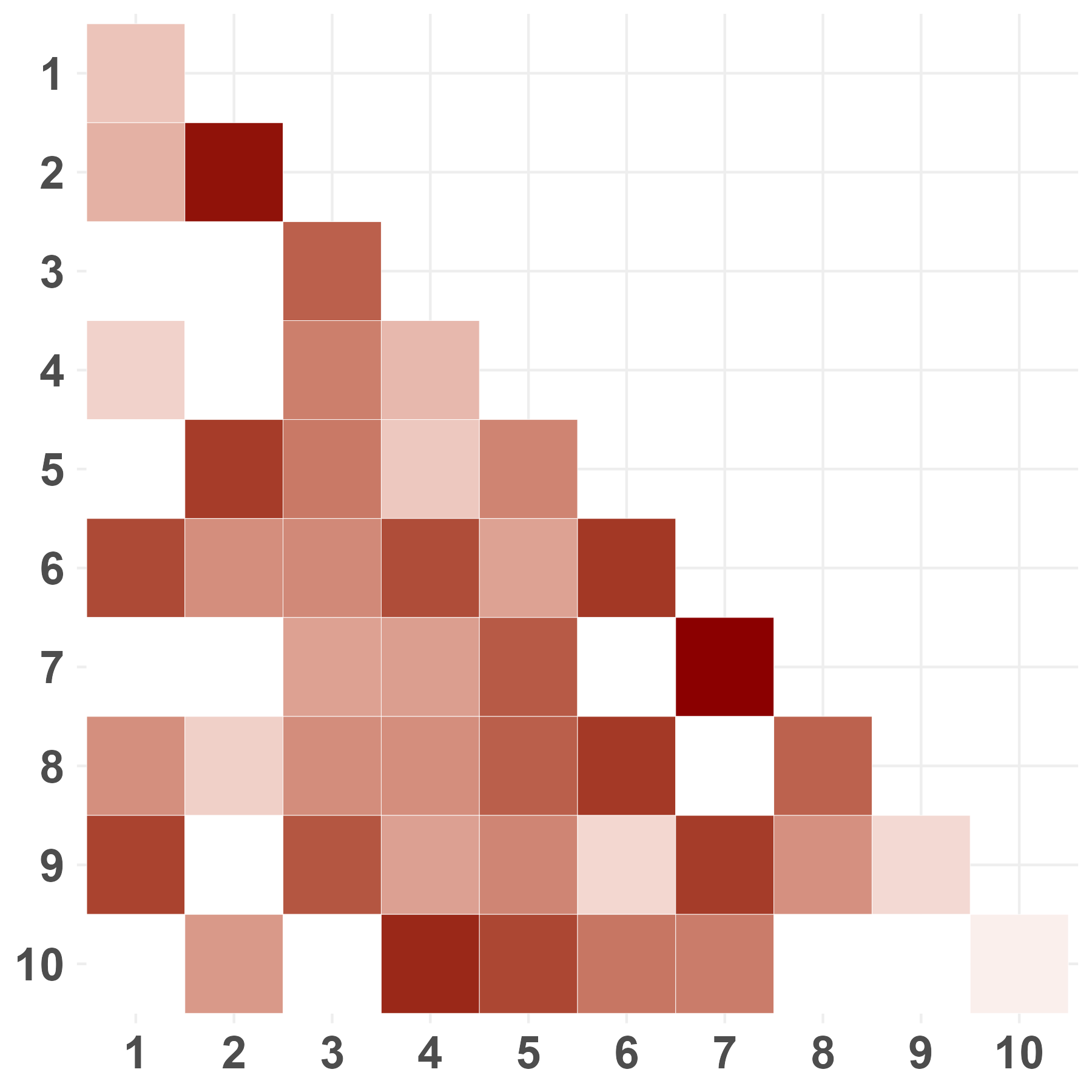}
         \includegraphics[width=0.48\textwidth]{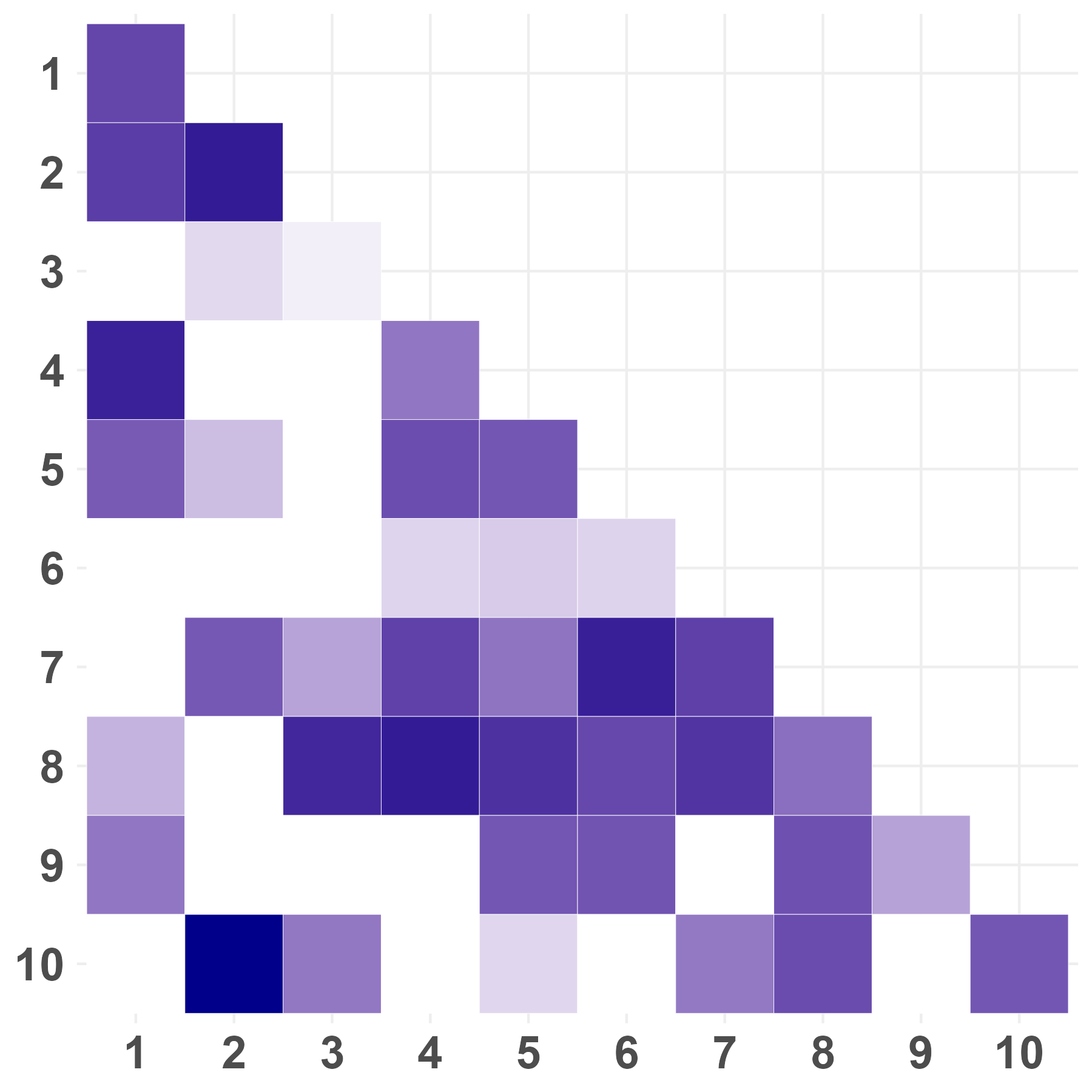}
         \caption{HCP-Emotion}
     \end{subfigure}
    
    \caption{Heatmaps summarize the number of the top $20\%$ selected connections with the highest absolute value of coefficient from CPM based on SBP(400) parcellation. The left and right column corresponds to the ABCD study and HCP study, respectively, and each row corresponds to a specific state. Within each subfigure, the connections with positive coefficients are colored red, and the negative coefficients are colored blue. The anonical neural networks in the plots correspond to: 1. medial frontal, 2. fronto parietal, 3. default mode, 4. motor, 5. visual I, 6. visual II, 7. visual association, 8. limbic,  9. basal ganglia, and 10. cerebellum. }
    \label{fig:abcd-hcp-cpm-plots}
\end{figure}

%% file: results/abcd-hcp-top5roi.tex
\begin{figure}[H]
\centering
    \begin{subfigure}{0.99\textwidth}
         \centering
         \includegraphics[width = 0.29\textwidth]{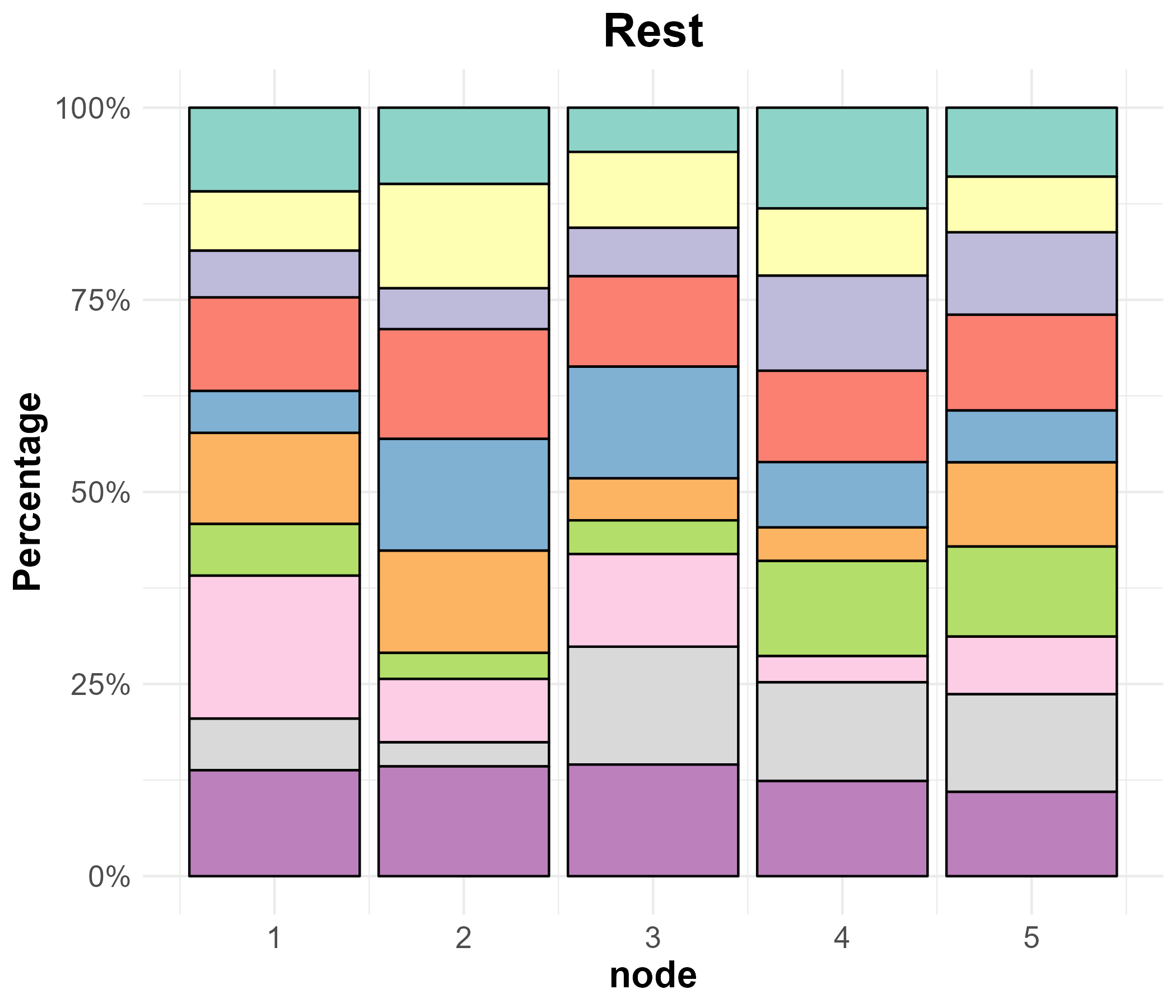}
         \includegraphics[width = 0.29\textwidth]{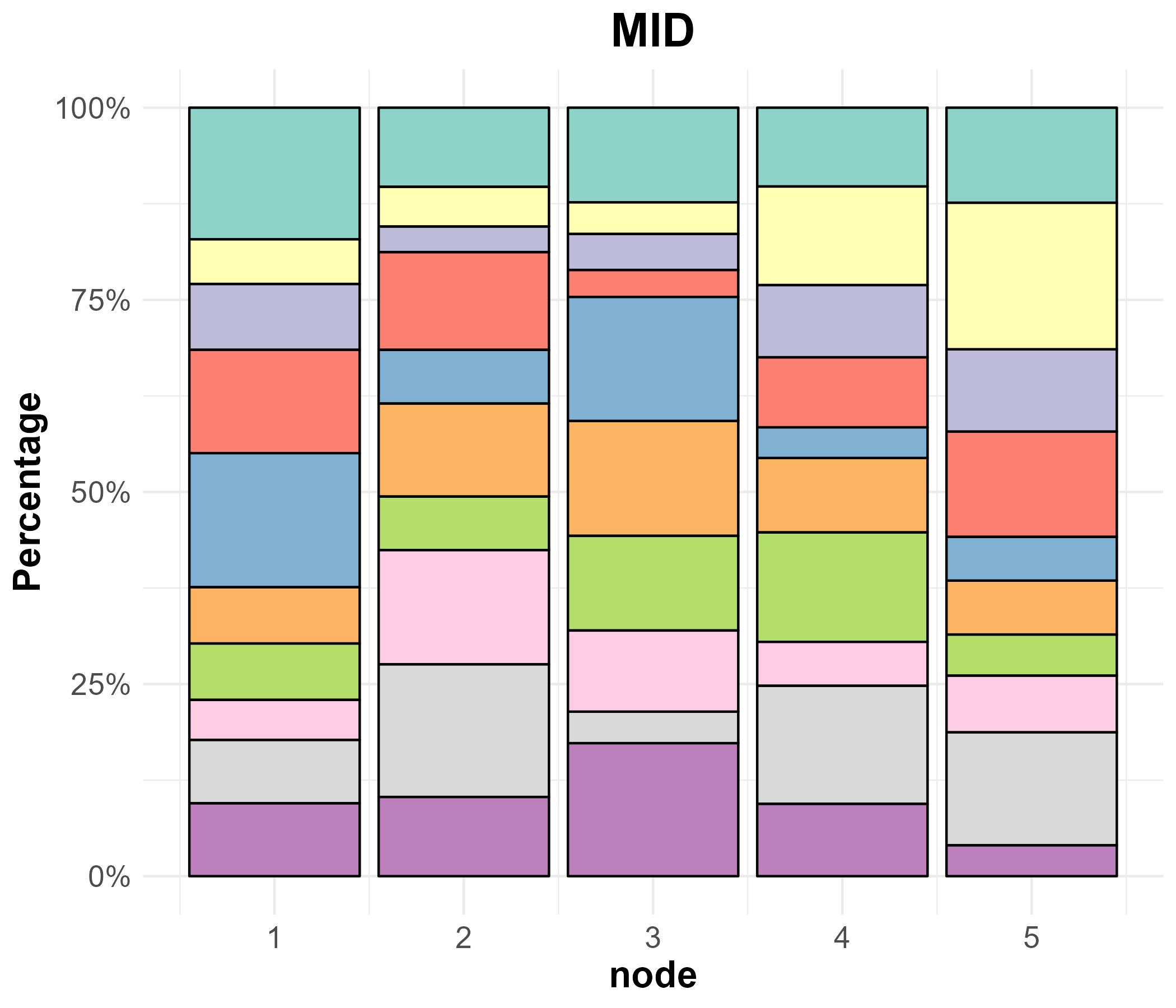}
         \includegraphics[width = 0.29\textwidth]{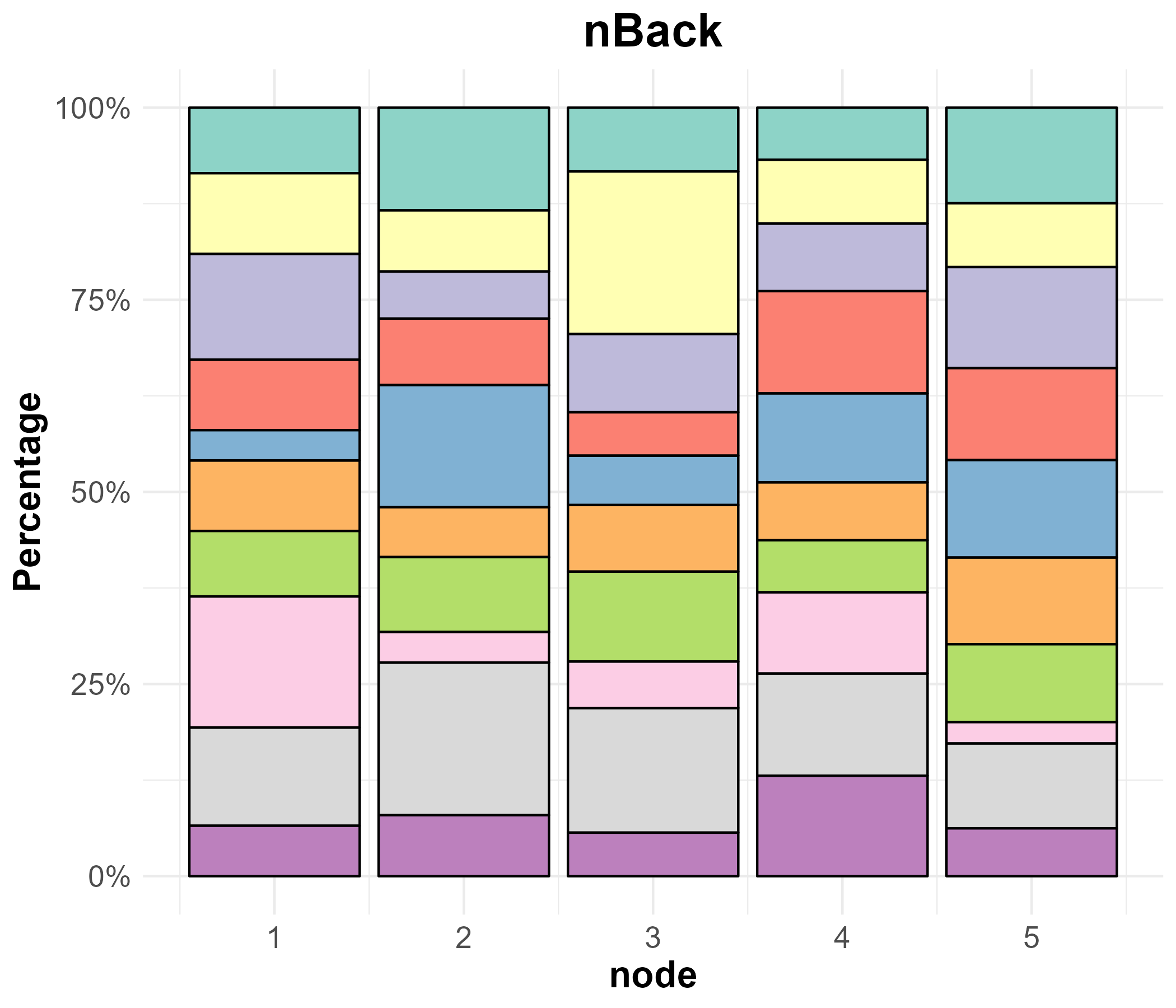}
        \includegraphics[height=4.1cm]{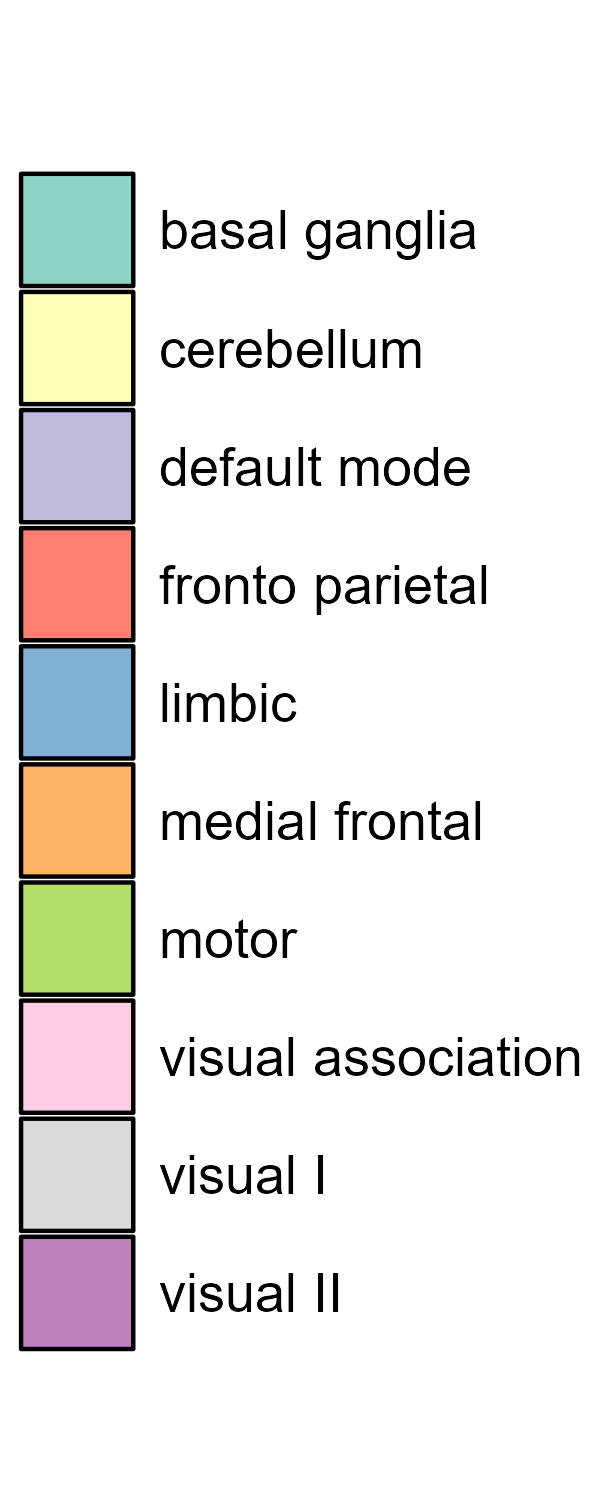}
          \caption{ABCD}
     \end{subfigure}
     \vfill
      \begin{subfigure}{0.99\textwidth}
         \centering
         \includegraphics[width = 0.29\textwidth]{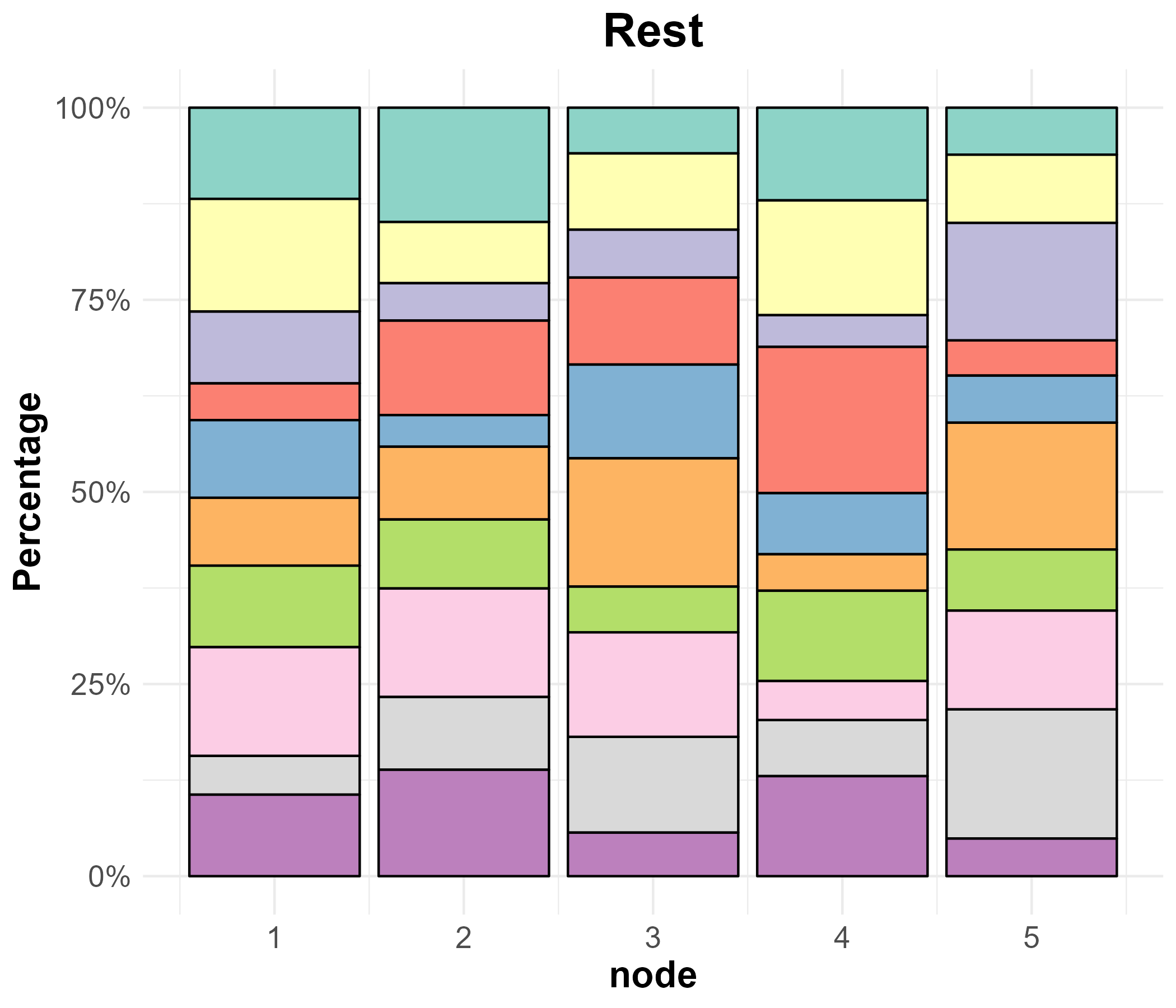}
         \includegraphics[width = 0.29\textwidth]{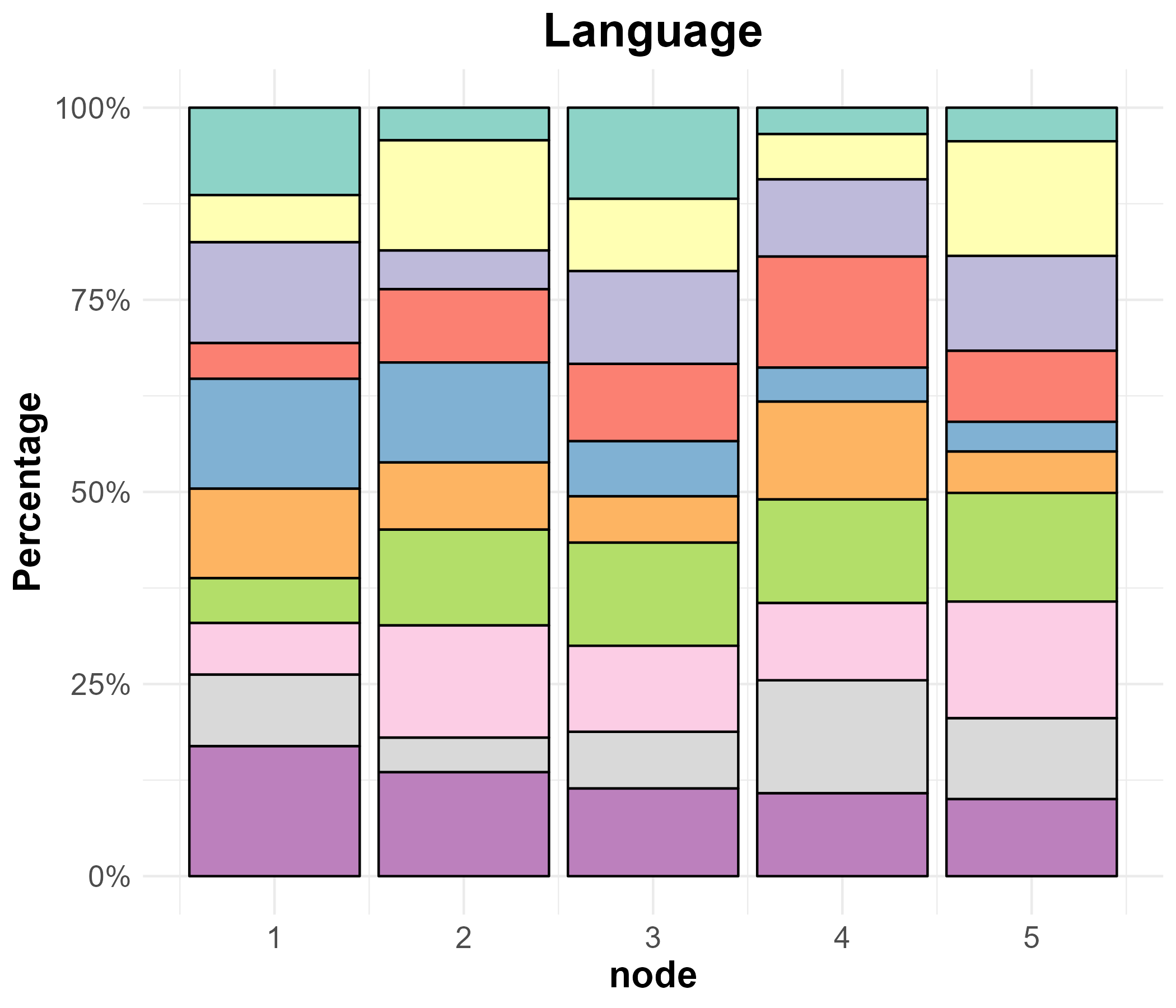}
         \includegraphics[width = 0.29\textwidth]{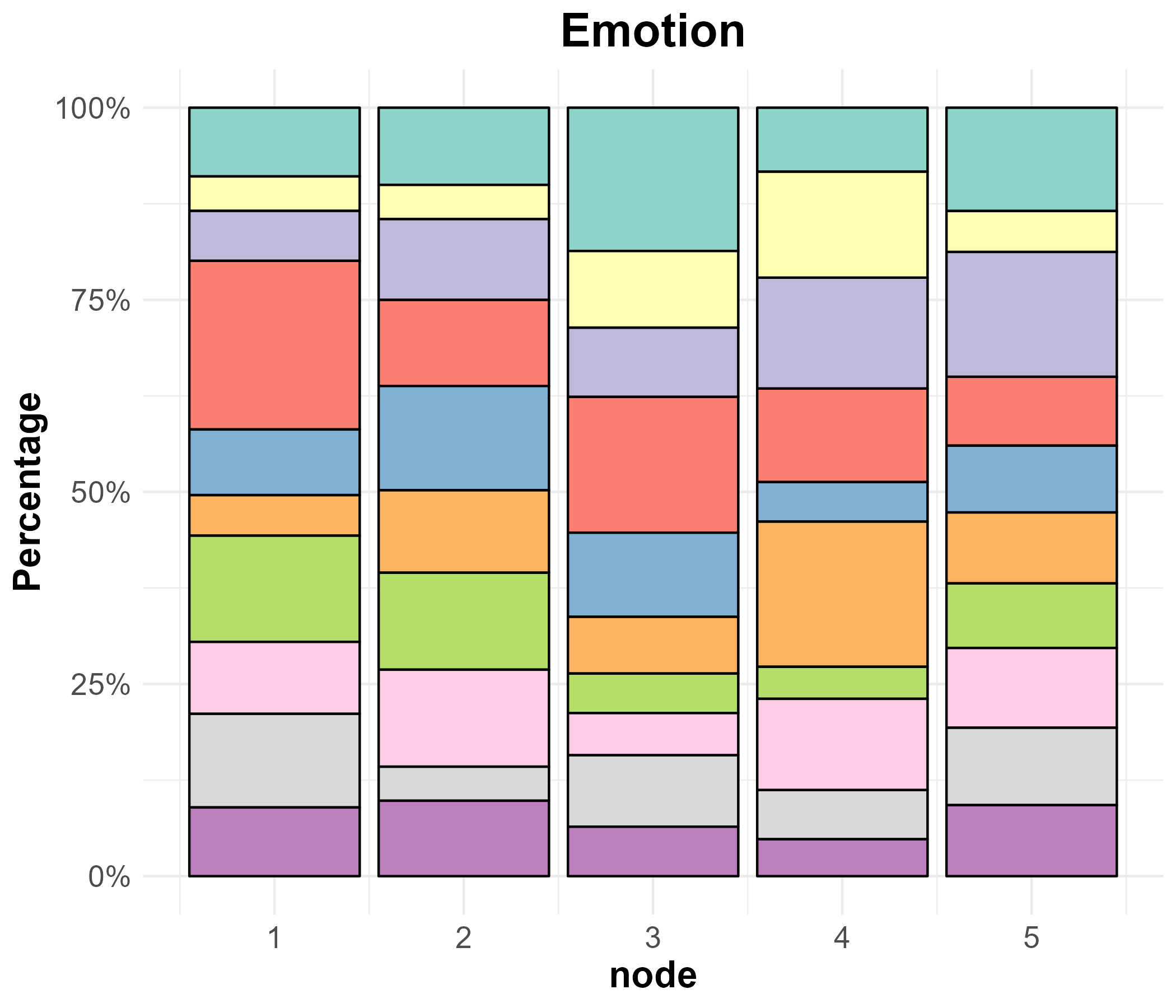}
          \includegraphics[height = 4.1cm]{imgs/top5roi-legend.png}
          \caption{HCP}
     \end{subfigure}
    \caption{Distribution among canonical neural networks for the largest five nodes constructed by SBP(400) for (a) ABCD study and (b) HCP study. }
    \label{fig:abcd-hcp-top5roi}
\end{figure}

%% file: results/abcd-hcp-ringplots.tex
\begin{figure}[H]
    \centering
     \begin{subfigure}[t]{0.49\textwidth}
         \centering
         Rest \hspace{+1.67cm} MID  \hspace{+1.67cm} nBack 
         \vfill
         \includegraphics[width=0.32\textwidth]{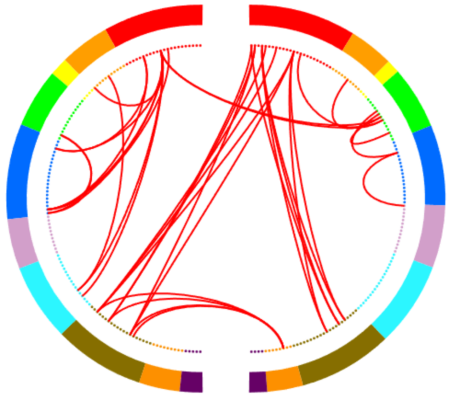}
         \includegraphics[width=0.32\textwidth]{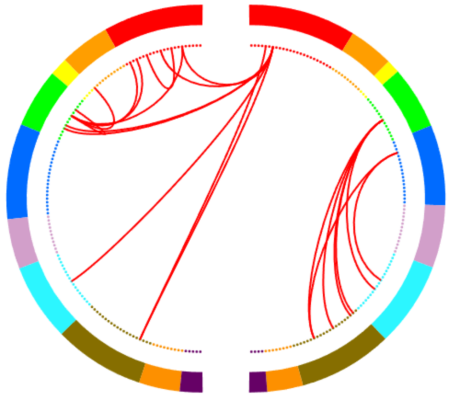}
         \includegraphics[width=0.32\textwidth]{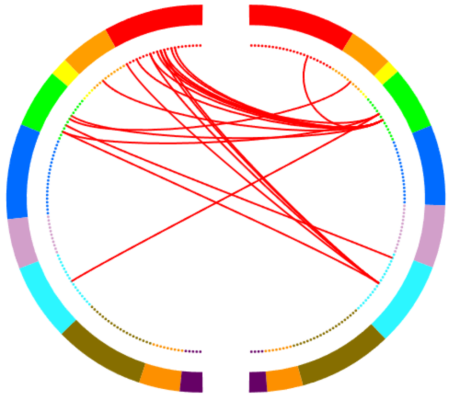}
         \includegraphics[width = 0.95\textwidth]{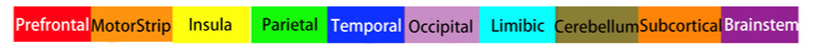}
         \caption{ABCD positive networks}
     \end{subfigure}
     \hfill
     \begin{subfigure}[t]{0.49\textwidth}
         \centering
         Rest \hspace{+1.66cm} MID  \hspace{+1.66cm} nBack 
         \vfill
         \includegraphics[width=0.32\textwidth]{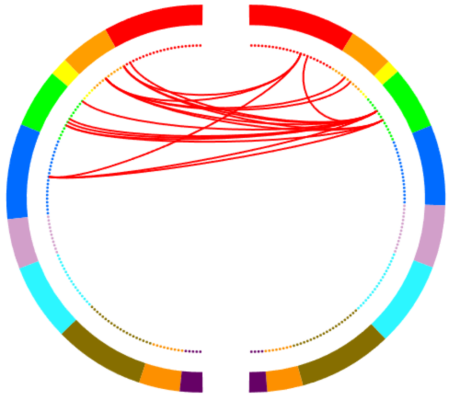}
         \includegraphics[width=0.32\textwidth]{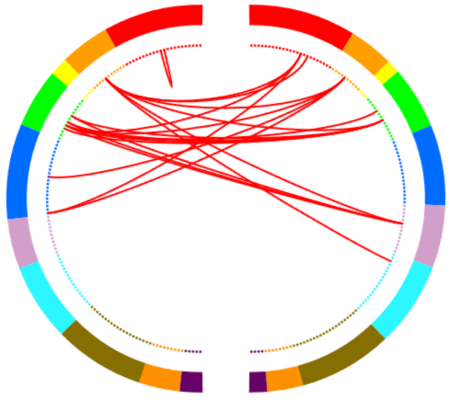}
         \includegraphics[width=0.32\textwidth]{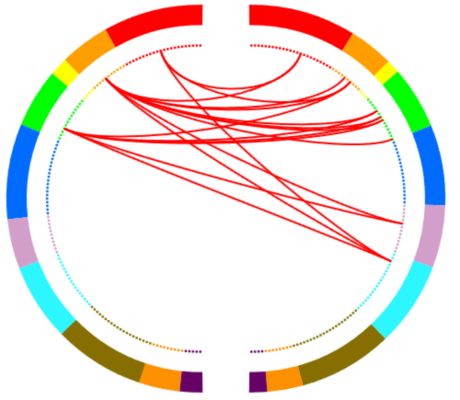}
         \includegraphics[width = 0.95\textwidth]{imgs/ringplot-legend-horizontal.png}
         \caption{HCP positive networks}
     \end{subfigure}
     \vfill
    \begin{subfigure}[t]{0.49\textwidth}
         \centering
         \includegraphics[width=0.32\textwidth]{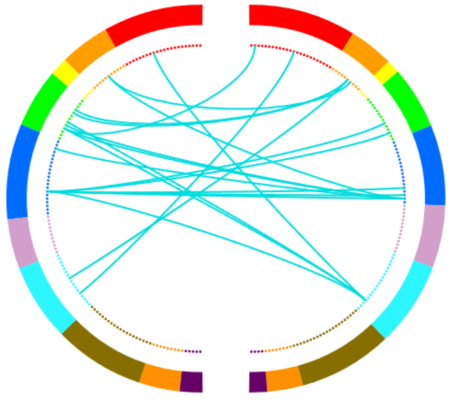}
         \includegraphics[width=0.32\textwidth]{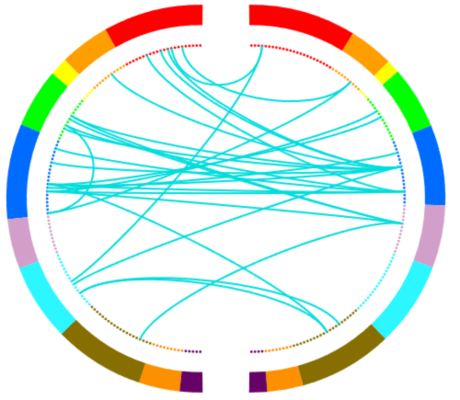}
         \includegraphics[width=0.32\textwidth]{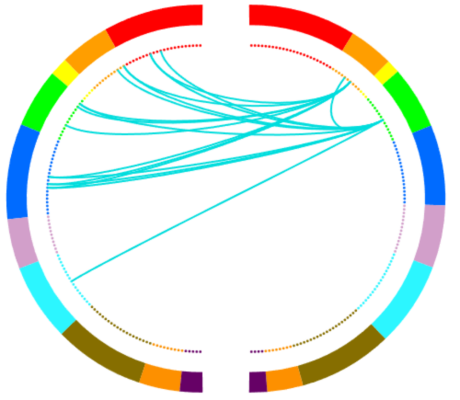}
         \includegraphics[width = 0.95\textwidth]{imgs/ringplot-legend-horizontal.png}
         \caption{ABCD negative networks}
     \end{subfigure}
     \hfill
     \begin{subfigure}[t]{0.49\textwidth}
         \centering
         \includegraphics[width=0.32\textwidth]{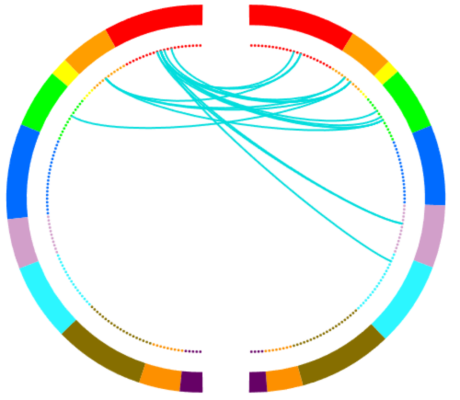}
         \includegraphics[width=0.32\textwidth]{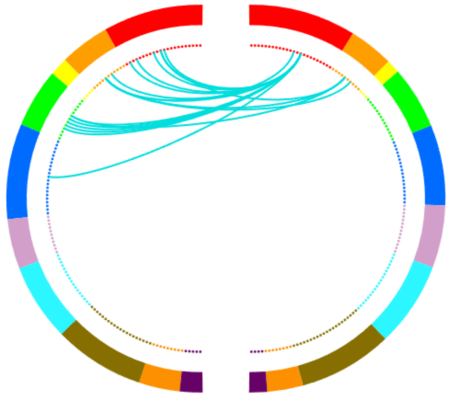}
         \includegraphics[width=0.32\textwidth]{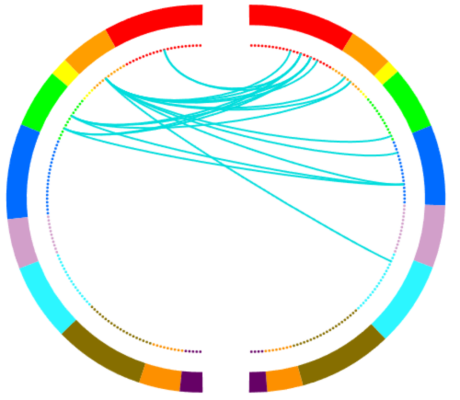}
         \includegraphics[width = 0.95\textwidth]{imgs/ringplot-legend-horizontal.png}
         \caption{HCP negative networks}
     \end{subfigure}
    \caption{A circular graph represents the significant positive and negative functional networks. Macroscale brain regions are color-coded as in the legend, and the cyan lines represent the significant connections. Subfigures (a), (b), (c) and (d) are the positive and negative networks for ABCD and HCP studies, respectively. }
    \label{fig:abcd-hcp-ringplots}
\end{figure}

%% file: results/abcd-hcp-reproducibility.tex
\begin{figure}[H]
    \centering
    \begin{subfigure}[t]{.9\textwidth}
    \centering
    \includegraphics[width = .49\textwidth]{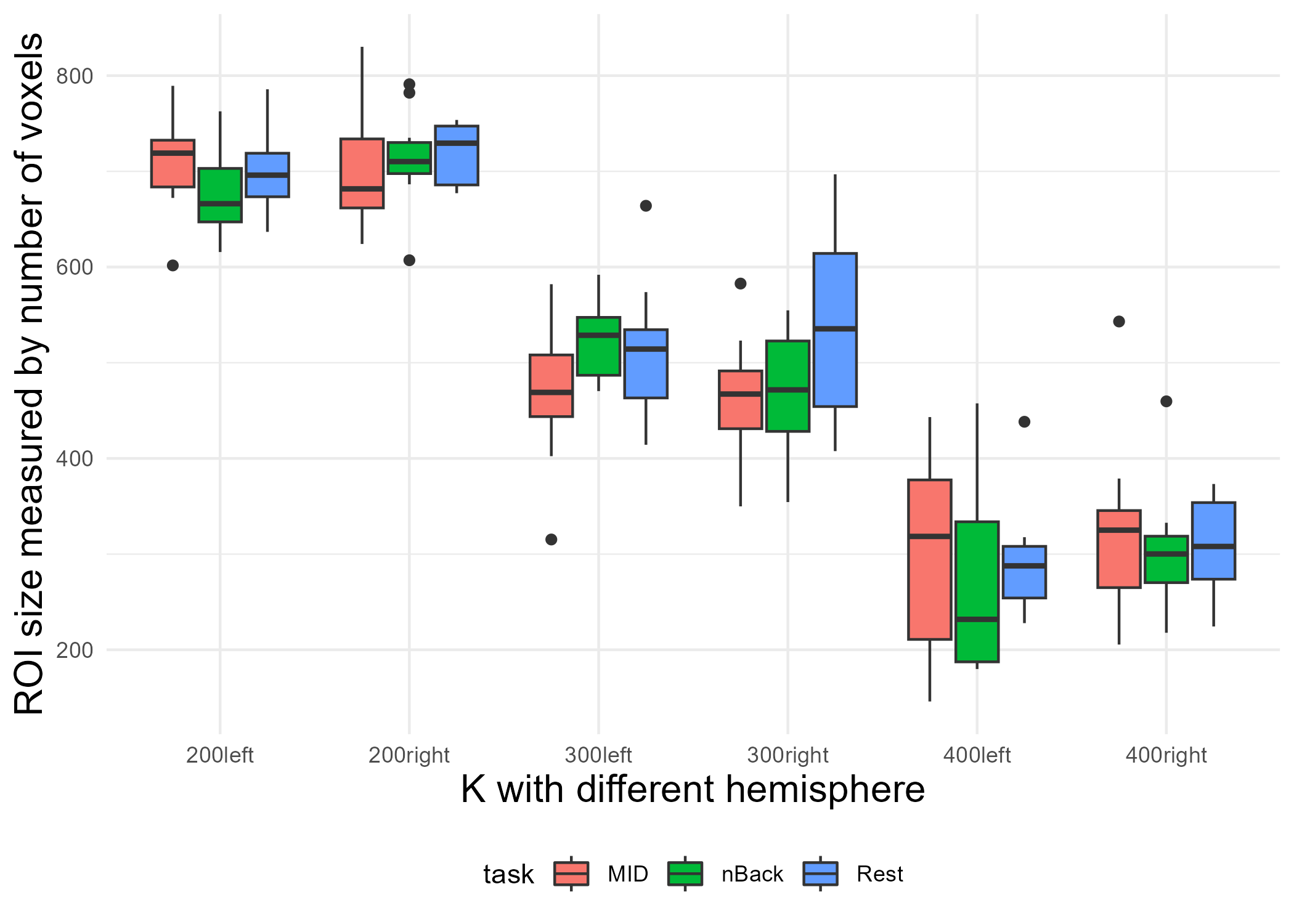}
    \includegraphics[width = .49\textwidth]{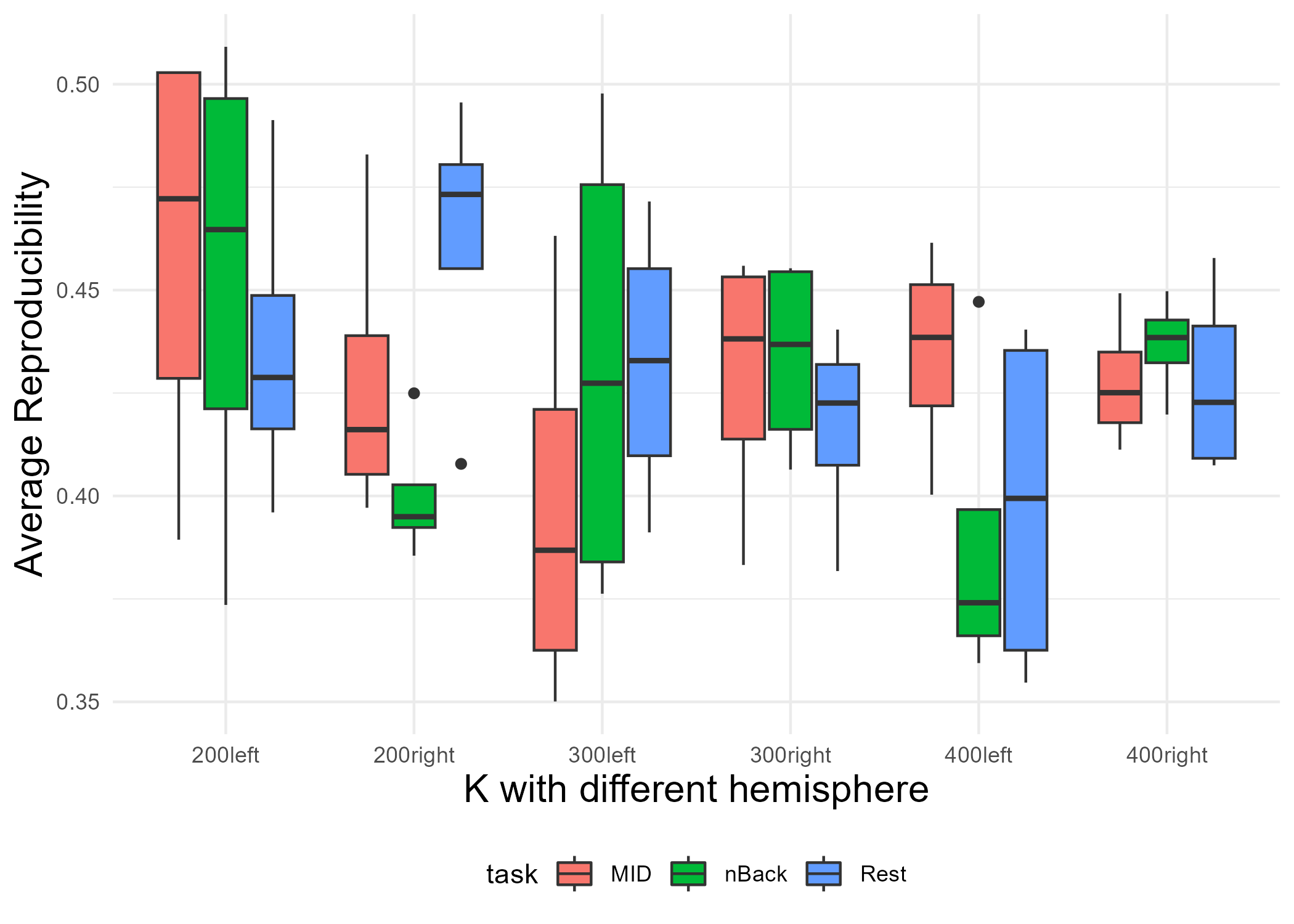}
    \caption{ABCD}
     \end{subfigure}
     \vfill
     
    \begin{subfigure}[t]{.9\textwidth}
    \centering
    \includegraphics[width =.49\textwidth]{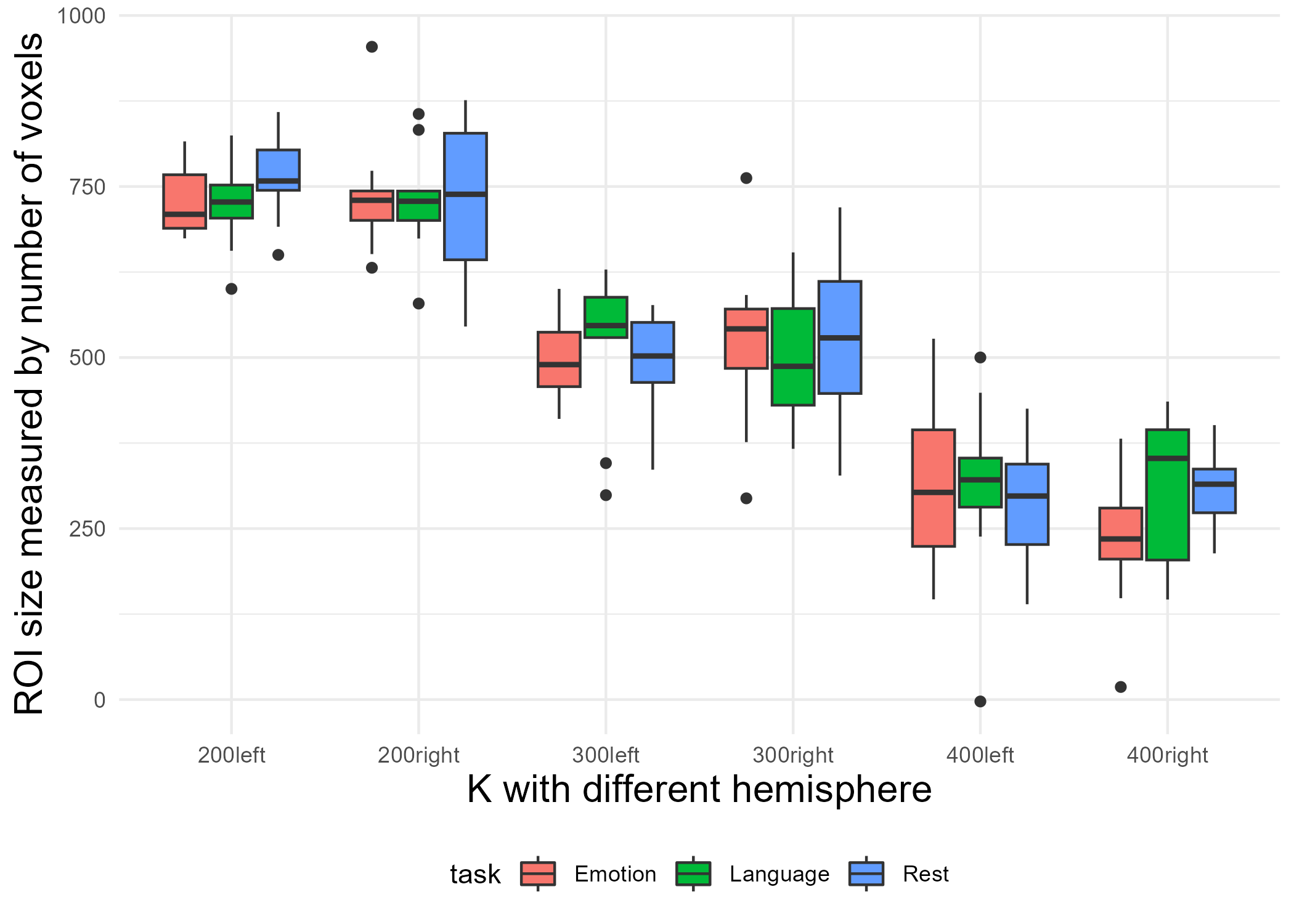}
    \includegraphics[width =.49\textwidth]{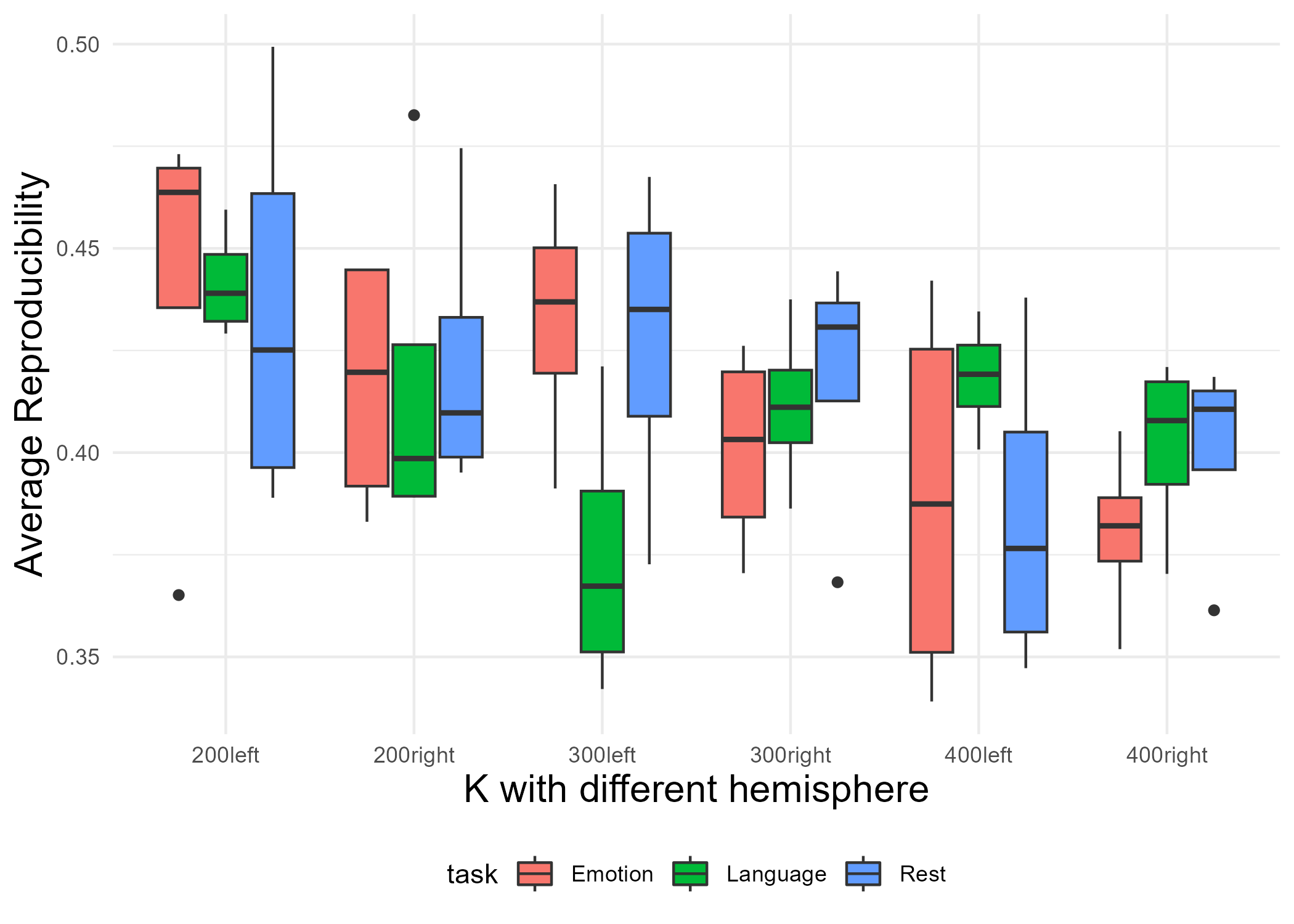}
    \caption{HCP}
     \end{subfigure}
    \caption{Node size stabilization (left) and average reproducibility score (right) for (a) ABCD and (b) HCP under different \( K \) and hemispheres. The task conditions are color-coded as in the legend. }
    \label{fig:abcd-hcp-reproducibility}
\end{figure}